\documentclass{IEEEtran}
\usepackage{amsmath,amssymb,amsfonts}
\usepackage{textcomp}
\def\BibTeX{{\rm B\kern-.05em{\sc i\kern-.025em b}\kern-.08em
T\kern-.1667em\lower.7ex\hbox{E}\kern-.125emX}}

\usepackage[pdftex]{graphicx}
\usepackage{caption}  
\usepackage{subcaption}
\graphicspath{ {./img/} }

\usepackage{cite}
\usepackage{makeidx}

\usepackage{algorithm}
\usepackage{algpseudocode}

\usepackage{multirow}
\usepackage{hyperref} 
\usepackage{cleveref} 

\hypersetup{colorlinks = true, 
	    linkcolor = black, 
	    urlcolor = black,
        citecolor = black} 

\begin{document}

\title{VBIM-Net: Variational Born Iterative Network \\ for Inverse Scattering Problems}

\author{\IEEEauthorblockN{Ziqing~Xing,~\IEEEmembership{Graduate~Student~Member,~IEEE, }
Zhaoyang~Zhang,~\IEEEmembership{Senior~Member,~IEEE, } \\
Zirui~Chen,~\IEEEmembership{Graduate~Student~Member,~IEEE, }
Yusong Wang,
Haoran Ma, \\
and Zhun Wei,~\IEEEmembership{Senior~Member,~IEEE} }     
\thanks{This work was supported in part by the National Natural Science Foundation of China under Grant 62394292 and Grant U20A20158, in part by Zhejiang Provincial Key Research and Development Program under Grant 2023C01021, in part by the Ministry of Industry and Information Technology under Grant TC220H07E, and in part by the Fundamental Research Funds for the Central Universities under Grant 226-2024-00069. 
(\textit{Corresponding author: Zhaoyang Zhang})
}
\thanks{Ziqing Xing and Zhaoyang Zhang are with the College of Information Science and Electronic Engineering, Zhejiang University, Hangzhou 310027, China, also with the Institute of Fundamental and Transdisciplinary Research, Zhejiang University, Hangzhou 310058, China, and also with Zhejiang Provincial Laboratory of Multi-Modal Communication Networks and Intelligent Information Processing, Hangzhou 310027, China (e-mail: ziqing\_xing@zju.edu.cn; ning\_ming@zju.edu.cn).

Zirui Chen is with the College of Information Science and Electronic Engineering, Zhejiang University, Hangzhou 310027, China, and also with the Zhejiang Provincial Laboratory of Multi-Modal Communication Networks and Intelligent Information Processing, Hangzhou 310027, China (e-mail: ziruichen@zju.edu.cn).

Yusong Wang and Zhun Wei are with the College of Information Science and Electronic Engineering, Zhejiang University, Hangzhou 310007, China (e-mail: 3170105272@zju.edu.cn; eleweiz@zju.edu.cn). 

Haoran Ma is with the School of Mathematical Sciences, Zhejiang University, Hangzhou 310007, China (e-mail: 12335002@zju.edu.cn).
}

}

\maketitle

\begin{abstract}
    Recently, studies have shown the potential of integrating field-type iterative methods with deep learning (DL) techniques in solving inverse scattering problems (ISPs). 
    In this article, we propose a novel Variational Born Iterative Network, namely, VBIM-Net, to solve the full-wave ISPs with significantly improved structural rationality and inversion quality. 
    The proposed VBIM-Net emulates the alternating updates of the total electric field and the contrast in the variational Born iterative method (VBIM) by multiple layers of subnetworks. 
    We embed the analytical calculation of the contrast variation into each subnetwork, converting the scattered field residual into an approximate contrast variation and then enhancing it by a U-Net, 
    thus avoiding the requirement of matched measurement dimension and grid resolution as in existing approaches. 
    The total field and contrast of each layer's output is supervised in the loss function of VBIM-Net, 
    imposing soft physical constraints on the variables in the subnetworks, which benefits the model's performance. 
    In addition, we design a training scheme with extra noise to enhance the model's stability. 
    Extensive numerical results on synthetic and experimental data both verify the inversion quality, generalization ability, and robustness of the proposed VBIM-Net. 
    This work may provide some new inspiration for the design of efficient field-type DL schemes.
\end{abstract}


\section{Introduction} \label{sec:introduction}
\IEEEPARstart{T}he inverse scattering problem (ISP) aims to determine the position, shape, and electromagnetic (EM) properties of unknown scatterers by measuring scattered fields \cite{chen2018computational}. 
This EM imaging technology has wide-ranging potential applications in geophysical exploration \cite{persico2014introduction, mangel2022multifrequency}, remote sensing \cite{kagiwada1990associate, tong2022environment}, security checks \cite{zhuge2010sparse, tan2020efficient}, biomedical imaging \cite{abubakar2002imaging, chandra2015opportunities}, integrated sensing and communication \cite{tong2021joint, tong2023multi}, etc. 
However, the inherent nonlinearity and ill-posedness of ISP make it challenging to design inversion algorithms. 

In order to obtain reliable inversion results, much effort has been devoted to addressing these two challenges, resulting in the iterative and non-iterative approaches, respectively. 
The non-iterative approaches are usually based on linear approximation or singular value decomposition (SVD), 
including Born approximation (BA) \cite{slaney1984Born}, Rytov approximation \cite{devaney1981Rytov}, back-propagation scheme (BPS) \cite{belkebir2005BPS}, and major current coefficient method \cite{yin2020MCC}, 
which are computationally economical but suffer from limited inversion accuracy. 
The iterative approaches formulate the optimization problems and find the optimal solution via iterations, 
which can be divided into field-type methods and source-type methods respectively according to updated physical quantities.  
Born iterative method (BIM) \cite{wang1989BIM} is a typical field-type method that applies BA in each iteration and alternately updates the contrast and the total field. Distorted Born iterative method (DBIM) \cite{chew1990DBIM} considers the variation of contrast as a disturbance to the inhomogeneous background, which improves the convergence of BIM but requires updating the Green's function during iterations. 
Variational Born iterative method (VBIM) \cite{zaiping2000VBIM} also retrieves the contrast variation and achieves the same convergence as DBIM, but it keeps Green's function unchanged to improve computational efficiency. 
Contrast source inversion (CSI) \cite{van1997CSI} and subspace optimization method (SOM) \cite{chen2009SOM} are typical source-type methods, which update the contrast and contrast source alternately using a two-step conjugate-gradient (CG) method. 
Some works \cite{ye2017DBIM_SOM,liu2019VBIM_SOM} enhanced the performance of field-type iterative methods by incorporating SOM. 
In addition, regularization techniques are often introduced to ease the ill-posedness of ISP \cite{vauhkonen1998tikhonov}, \cite{van1995total}, \cite{shea2011tsvd}. 
These iterative methods are able to obtain relatively high-quality inversion results, 
but the usually high computational overhead makes them difficult to perform real-time reconstruction. 

Recently, researchers have successfully applied deep learning (DL) to solve ISPs. 
Compared with traditional approaches, DL has significant advantages in reconstruction quality and computational efficiency. 
The first type of DL-based ISP algorithm combines the low-accuracy inversion solver with the deep neural networks (DNNs). 
Usually, these methods use BPS to obtain a rough permittivity image and then enhance it through a image-to-image translation network such as U-Net \cite{ronneberger2015UNet} and generative adversarial network (GAN) \cite{goodfellow2014GAN}.
Wei and Chen \cite{wei2018DLS} proposed a dominant current scheme to replace BPS, aiming to generate a better input image for U-Net. 
Song et al.  \cite{song2021GAN} designed a novel perceptual GAN to improve the image-to-image translation.
Wei and Chen \cite{wei2019ICLM} introduced a induced current learning scheme to perform image translation in the contrast source domain instead of permittivity.
Wang et al. \cite{wang2023MSDLS} combined real space and frequency space information in parallel and serial ways to enhance the reconstruction quality of U-Net.  
Even though these methods yield satisfactory inversion results, they still lack physical insights, only incorporating physical information in the rough reconstruction stage, which limits their generalization ability.

The second type of DL-based algorithm integrates traditional iterative methods with DNNs to introduce physical information into network design. 
Sanghvi et al. \cite{sanghvi2019CSNet} designed contrast source net (CS-Net) to learn the noise space component of the contrast source current and embedded it into two-fold SOM \cite{zhong2009TSOM}, which has a satisfying generalization ability but is still time-consuming for the iterative process. 
Liu et al. \cite{liu2021PMNet} transformed the ISP into a constrained optimization problem and designed a physical model-inspired deep unrolling network (PM-Net) to emulate the alternating direction method of multiplier (ADMM) iterative process, updating the contrast source and contrast alternately. 
Zhang et al. \cite{zhang2022CSINet} embedded CSI method into a convolutional neural network (CNN) structure. 
Liu et al. \cite{liu2022SOMNet} proposed SOM-Net by replacing the contrast source update step in SOM iteration with a U-Net. 
In general, the aforementioned schemes all combine source-type iterative methods with DNN design, which achieve better generalization than typical image-to-image translation schemes. 
In parallel, there has been a few works exploring the potential of integrating DNNs with field-type iterative methods, treating the contrast and the total field as inversion variables. 
Beerappa et al. \cite{beerappa2023deep} embedded two cascaded CNNs into the iteration of DBIM, which reduces the reconstruction error of DBIM, but still demands heavy computation within iterations. 
In \cite{shan2022NeuralBIM}, Shan et al. proposed the neural Born iterative method (NeuralBIM) by utilizing the physics-informed supervised residual learning (PhiSRL) \cite{shan2023PhiSRL} to fully unroll the BIM iteration. 
However, NeuralBIM directly concatenates the scattered field residual and the contrast image as the network input, which in general requires the measurement dimension to well match the resolution of the inversion grid. 
More importantly, although NeuralBIM attempts to use multiple layers of subnetworks with the same structure to emulate the iterations, the computations and outputs of each intermediate layer are not governed by any form of physical constraints. 
As a result, its data flow fails to well conform to that of the original iterations, thus degrading the model's physical interpretability and generalization ability. 

In this paper, we propose a novel variational Born iterative network, namely, VBIM-Net, by unrolling the VBIM algorithm to solve the full-wave ISPs. 
We embed physics-inspired analytical computations into the network architecture and incorporate soft physical constraints into the loss function, aiming to overcome the drawbacks of the existing field-type DL schemes in terms of structural rationality and inversion quality. 
In each subnetwork, we transform the scattered field residual into an approximate contrast variation inspired by VBIM, 
which naturally lifts the limitation on measurement dimension and gird resolution in NeuralBIM without sacrificing computational efficiency. 
Through the proposed layer-wise constraint, the predicted contrast and total field in each subnetwork are explicitly supervised in the loss function, guiding the data flow to conform to an iterative process and thereby enhancing the model's reliability and interpretability. 
In the training stage, we adopt a hybrid dataset and design a training scheme with extra noise further to ensure the generalization ability and robustness of the model. 
Synthetic and experimental experiments both verify the superior performance of the proposed VBIM-Net over the existing ones. 
The main contributions of this work are summarized as follows.
\begin{enumerate}
    \item The proposed VBIM-Net transforms the scattered field residual into approximate contrast variation through analytical computation. 
    It aligns the input variables of the contrast update network in spatial dimensions and unifies them into the contrast domain. 
    This avoids the requirement for matched measurement dimension and grid resolution in NeuralBIM, while fully utilizing the information from the scattered field residual to update the contrast, 
    which improves the structural rationality of the existing field-type DL scheme. 
    \item A loss function with a layer-wise constraint is designed, 
    ensuring the outputs of each VBIM-Net layer to be consistent with the actual physical quantities, 
    while providing more accurate physical quantity residuals for the next layer. 
    This makes the data flow of VBIM-Net closer to a iterative algorithm, thereby ensuring the reliability and generalization performance when training the deep VBIM-Net model. 
    \item In the training stage, we mix two synthetic datasets to enable the model to learn the inversion of homogeneous and inhomogeneous scatterers. 
    Besides, we implement a training scheme with extra noise and employ a loss function that scales with the signal-to-noise ratio (SNR), thereby enhancing the robustness of the model. 
    \item Extensive experiments have been performed on synthetic and experimental datasets, including challenging profiles with complex shapes, high contrast, and high noise level. 
    The proposed VBIM-Net demonstrates significant advantages in terms of reconstruction quality, generalization ability, and robustness. 
    We also conduct a series of experiments to verify the effectiveness and necessity of the novel designs proposed in VBIM-Net. 
\end{enumerate}

This article is organized as follows. 
Section \ref{sec:problem_and_method} formulates ISPs and introduces BPS, BIM, and VBIM. 
In Section \ref{sec:net}, we introduce the design of the proposed VBIM-Net and analyze its computational complexity. 
Section \ref{sec:numerical_results} shows experimental results on synthetic and experimental data for the performance verification. 
Finally, we conclude our work in Section \ref{sec:conclusion}.

\section{Problem Formulation and Basic Methods} \label{sec:problem_and_method}
\begin{figure} [!htbp] 
\centering
\includegraphics[width=0.8\linewidth]{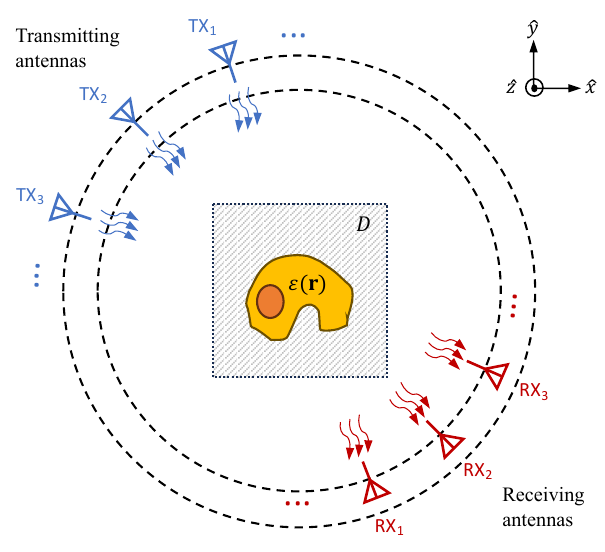}
\captionsetup{font=footnotesize,justification=raggedright,singlelinecheck=false}
\caption{The schematic of a 2-D ISP under TM illuminations, where targets are inside the DOI $D$, and transmitting and receiving antennas are located on the circumference.}
\label{fig:scenerio}
\end{figure}

For convenience of presentation, we consider the 2-D transverse-magnetic (TM) case, where the longitude direction is along $\hat{z}$.
As shown in Fig. \ref{fig:scenerio}, the dielectric scatterers are placed in a domain of interest (DOI) $D\subset R^2$, which individually illuminated by $N_i$ transmitters (TXs) located at $\mathbf{r}^i_p, p = 1,\cdots,N_i$. 
For each illumination, the scattered field is measured by $N_r$ receivers (RXs) located at $\mathbf{r}^r_q, q = 1,\cdots,N_r$.

\subsection{Problem Formulation} \label{subsec:problem_formulation}

We use $E^i(\mathbf{r}), E^t(\mathbf{r})$, and $E^s(\mathbf{r})$ to denote the incident field, the total field, and the scattered field at location $\mathbf{r}$, respectively.
The $E^t$ generated by $E^i$ is determined by the Lippmann–Schwinger equation 
\begin{equation} \label{eq:continue_state_eq}
    E^t(\mathbf{r}) = E^{i}(\mathbf{r}) + k_0^2 \int_{D} g(\mathbf{r},\mathbf{r}^{\prime}) \chi(\mathbf{r}^{\prime}) E^t(\mathbf{r}^{\prime})d\mathbf{r}^{\prime}, \enspace \text{for }\mathbf{r}\in D, 
\end{equation}
where $k_0$ is the wavenumber of the free space, and $g(\mathbf{r},\mathbf{r}^{\prime})$ is the scalar 2-D Green’s function.
The contrast is defined as $\chi(\mathbf{r}) = \epsilon_{r}(\mathbf{r})-1$ with the relative permittivity $\epsilon_{r}(\mathbf{r})$.
The contrast source is derived by $J(\mathbf{r})=\chi(\mathbf{r})E^t(\mathbf{r})$.
The scattered field is the re-radiation of the contrast source, so the $E^s$ measured by the $q$th receiving antenna is
\begin{equation} \label{eq:continue_data_eq}
    E^s(\mathbf{r}^{r}_q) = k_0^2 \int_{D} g(\mathbf{r}^{r}_q,\mathbf{r}^{\prime}) \chi(\mathbf{r}^{\prime}) E^t(\mathbf{r}^{\prime})d\mathbf{r}^{\prime}.
\end{equation}

Referring to \cite{chen2018computational}, we divide $D$ into an $M\times M$ grid, using $\mathbf{r}_n$ to denote the center of the $n$th subunit with $n=1,\cdots,M^2$. The area of each subunit is $S$. 
By applying the method of moment (MoM), we can obtain the discretized forms of Eq. (\ref{eq:continue_state_eq}) and (\ref{eq:continue_data_eq}) as follows, 
\begin{align}
    \mathbf{E}^{t} &= \mathbf{E}^{i} + \mathbf{G}_{D}\operatorname{diag}(\boldsymbol{\chi}) \mathbf{E}^{t}, \label{eq:discrete_state_eq}  \\
    \mathbf{E}^{s} &= \mathbf{G}_{S}\operatorname{diag}(\boldsymbol{\chi}) \mathbf{E}^{t},  \label{eq:discrete_data_eq} 
\end{align}
where $\mathbf{E}^{i}, \mathbf{E}^{t}\in\mathbb{C}^{M^2\times N_i}$, $\mathbf{E}^{s}\in\mathbb{C}^{N_r\times N_i}$. 
The $\boldsymbol{\chi}\in\mathbb{C}^{M^2\times 1}$ if the targets are lossy scatters, and $\boldsymbol{\chi}\in\mathbb{R}^{M^2\times 1}$ otherwise. 
$\operatorname{diag}(\boldsymbol{\chi})$ represents the diagonal matrix composed of elements from vector $\boldsymbol{\chi}$. 
The (\ref{eq:discrete_state_eq}) is called the state equation, and the (\ref{eq:discrete_data_eq}) is called the data equation. 
In the following, we use $\mathbf{E}^{i}_p, \mathbf{E}^{t}_p$, and $\mathbf{E}^{s}_p$ to represent the incident field, total field, and scattered field under the $p$th incidence, which are the $p$th columns of $\mathbf{E}^i$, $\mathbf{E}^t$, and $\mathbf{E}^s$, respectively.
The $n$th elements of $\mathbf{E}^{i}_p, \mathbf{E}^{t}_p$, and $\boldsymbol{\chi}$ represent the $E^{i}, E^{t}$, and $\chi$ at $\mathbf{r}_n$, 
and the $\mathbf{E}^{s}_p$ denotes the scattered field measured at each receiving antenna. 
The $\mathbf{G}_{D} \in \mathbb{C}^{M^2\times M^2}$ and $\mathbf{G}_{S} \in \mathbb{C}^{N_r\times M^2}$ are discrete forms of the Green's function,
\begin{align}
    \mathbf{G}_{D}[n, n^{\prime}] &= \begin{cases}\begin{aligned}
        &\dfrac{jk_{0}\pi a}{2} J_{1}(k_{0}a)H_{0}^{(1)}(k_{0}|\mathbf{r}_{n}-\mathbf{r}_{n^{\prime}}|), &&\text{if } n\neq n^{\prime} \\
        &\dfrac{jk_{0}\pi a}{2} H_{1}^{(1)}(k_{0}a)-1, &&\text{if } n= n^{\prime}
    \end{aligned}\end{cases}\\
    \mathbf{G}_{S}[q, n] &= \frac{jk_{0}\pi a}{2} J_{1}(k_{0}a)H_{0}^{(1)}(k_{0}|\mathbf{r}^{r}_{q}-\mathbf{r}_{n}|),
\end{align}
where each square subunit is approximated as a cicle with the same area, which has an equivalent radius $a=\sqrt{S/\pi}$.
The $H_0^{(1)}(\cdot), H_1^{(1)}(\cdot)$, and $J_1(\cdot)$ denote the zeroth order, the first order of Hankel function of the first kind, and the Bessel function of the first order, respectively.

The forward problem is to compute the scattered field $\mathbf{E}^s$ using the incident field $\mathbf{E}^i$ and the contrast $\boldsymbol{\chi}$ of targets according to Eq. (\ref{eq:discrete_state_eq}) and (\ref{eq:discrete_data_eq}).
Correspondingly, ISPs aim to solve the $\boldsymbol{\chi}$ based on the $\mathbf{E}^i$ and $\mathbf{E}^s$.

\subsection{Back-Propagation Scheme} \label{subsec:BPS}
BPS is an efficient noniterative inversion algorithm that can provide initial solutions for iterative algorithms \cite{van1997CSI, chen2009SOM} and neural networks \cite{wei2018DLS,wang2023MSDLS}. 
BPS models the contrast source as proportional to the backpropagated field, which is effective in the weak scattering case \cite{chen2018computational}. 
For the $p$th incidence, the solution of contrast source is
\begin{equation} \label{eq:BPS_ICC}
    \mathbf{J}^b_p = \gamma_p \cdot \mathbf{G}_{S}^H \mathbf{E}^s_p,
\end{equation}
where the coefficient $\gamma_p$ is 
\begin{equation}
    \gamma_p = \frac{(\mathbf{E}^s_p)^T (\mathbf{G}_{S} \mathbf{G}_{S}^H \mathbf{E}^s_p)^*}{\left\|\mathbf{G}_{S} \mathbf{G}_{S}^H \mathbf{E}^s_p\right\|^2}.
\end{equation}
With Eq. (\ref{eq:discrete_state_eq}), the total field of the $p$th incidence is 
\begin{equation}
    \mathbf{E}^{t,b}_p = \mathbf{E}^{i}_p + \mathbf{G}_{D} \mathbf{J}^{b}_p. 
\end{equation}
According to the definition of contrast source, 
\begin{equation}
    \mathbf{J}_p^b=\operatorname{diag}(\boldsymbol{\chi}^b) \mathbf{E}_p^{t,b},
\end{equation}
the contrast $\boldsymbol{\chi}^b$ can be obtained by the least square (LS) method, and its $n$th element is
\begin{equation} \label{eq:BPS_x}
    \boldsymbol{\chi}^b(n)=\frac{\sum_{p=1}^{N_i} \mathbf{J}_p^b(n) \left[\mathbf{E}_p^{t,b}(n)\right]^*}{\sum_{p=1}^{N_i}\left\|\mathbf{E}_p^{t,b}(n)\right\|^2}.
\end{equation}

\subsection{Born Iterative Method and Its Variational Version} \label{subsec:BIM}
The BIM \cite{wang1989BIM} is a classic iterative inversion algorithm that reconstructs scatterers by alternately updating the contrast $\boldsymbol{\chi}$ and the total field $\mathbf{E}^t$.
The initial value of contrast $\boldsymbol{\chi}_{(0)}$ can be obtained using the BPS. 
For each iteration step $k$, the total field $\mathbf{E}^t$ is calculated according to the Eq. (\ref{eq:discrete_state_eq}), 
\begin{equation} \label{eq:Et_update_BIM}
    \mathbf{E}^{t}_{(k)} = \left(\mathbf{I} - \mathbf{G}_{D} \operatorname{diag}\left(\boldsymbol{\chi}_{(k-1)}\right)\right)^{-1} \mathbf{E}^i.
\end{equation}
Then, BIM updates the contrast $\boldsymbol{\chi}$ by Born approximation for homogeneous background according to Eq. (\ref{eq:discrete_data_eq}), 
\begin{equation} \label{eq:data_eq_linear_approx}
    \mathbf{E}^s = \mathbf{G}_{S} \operatorname{diag}(\boldsymbol{\chi}) \mathbf{E}^{t}_{(k)}, 
\end{equation}
which can be reduced to a LS estimation in noisy conditions. 
To alleviate the ill-posedness of ISPs, L2 regularization is commonly introduced, which can be formulated as
\begin{equation} \label{eq:x_update_BIM}
    \boldsymbol{\chi}_{(k)} = \arg\min_{\boldsymbol{\chi}} {\left\{\|\mathbf{E}^s - \mathbf{G}_{S} \operatorname{diag}(\boldsymbol{\chi}) \mathbf{E}^{t}_{(k)}\|_2^2 + \lambda\cdot \|\boldsymbol{\chi}\|_2^2\right\}} .
\end{equation}

As an evolved version of BIM, VBIM adopts the variational form of Eq. (\ref{eq:data_eq_linear_approx}) in the $\boldsymbol{\chi}$-update step, 
\begin{equation} \label{eq:delta_Es}
    \delta\mathbf{E}^s_{(k)} = \mathbf{G}_{S} \operatorname{diag}(\delta\boldsymbol{\chi}) \mathbf{E}^{t}_{(k)}, 
\end{equation}
where $\delta\mathbf{E}^s_{(k)} = \mathbf{E}^s - \mathbf{E}^s_{(k)}$ is the residual of scattered field, 
$\mathbf{E}^s_{(k)}$ is the computed scattered field in the $k$th iteration, 
\begin{equation} \label{eq:Es_k}
    \mathbf{E}^s_{(k)} = \mathbf{G}_{S} \operatorname{diag}\left(\boldsymbol{\chi}_{(k-1)}\right) \mathbf{E}^{t}_{(k)}, 
\end{equation}
and $\delta\boldsymbol{\chi} = \boldsymbol{\chi}_{(k)} - \boldsymbol{\chi}_{(k-1)}$ is the difference between the contrast value in the $k$th and $(k-1)$th iteration.
Compared with BIM, VBIM has faster convergence and better inversion quality.

\section{The Proposed Variational Born Iterative Network} \label{sec:net}
To fully leverage the nonlinear fitting capability of DNNs and avoid the computational overhead from iterative calculations, 
the proposed VBIM-Net simulates the VBIM iterations by multiple layers of subnetworks with the same structure.  
Each subnetwork embeds the contrast variation in VBIM to transfer the information from scattered field residual.
In order to make the variables in each subnetwork correspond to physical quantities, we introduce a layer-wise constraint in the loss function, which ensures the model's reliability. 
Furthermore, we employ a training scheme with extra noise to enhance the robustness of the model. 

In this section, we will introduce the specific design of VBIM-Net in network architecture and loss function, and analyze its computational complexity.

\subsection{Network Architecture} \label{subsec:VBI_net}

\begin{figure*}[!t]
    \centering
    \includegraphics[width=\linewidth]{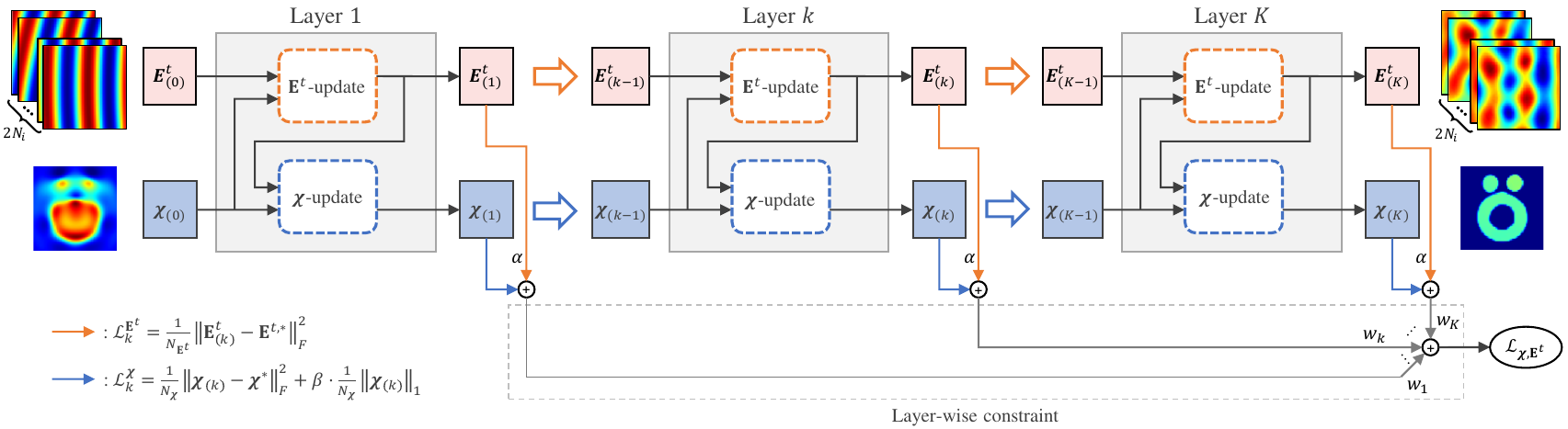}
    \captionsetup{font=footnotesize}
    \caption{The framework of the proposed VBIM-Net. 
    VBIM-Net is composed of $K$-layer subnetworks with the same structure but independent parameters. 
    The input of VBIM-Net is the incident field of $N_i$ incidence and the rough contrast image obtained by BPS. 
    The output of each VBIM-Net layer includes the predicted total field $\mathbf{E}^t_{(k)}$ and contrast $\boldsymbol{\chi}_{(k)}$, $k = 1, \cdots, K$. 
    The loss function calculation is depicted under the network framework. 
    }
    \label{fig:VBIM_Net_and_loss}
\end{figure*}

\begin{figure*}[!t]
    \centering
    \includegraphics[width=\linewidth]{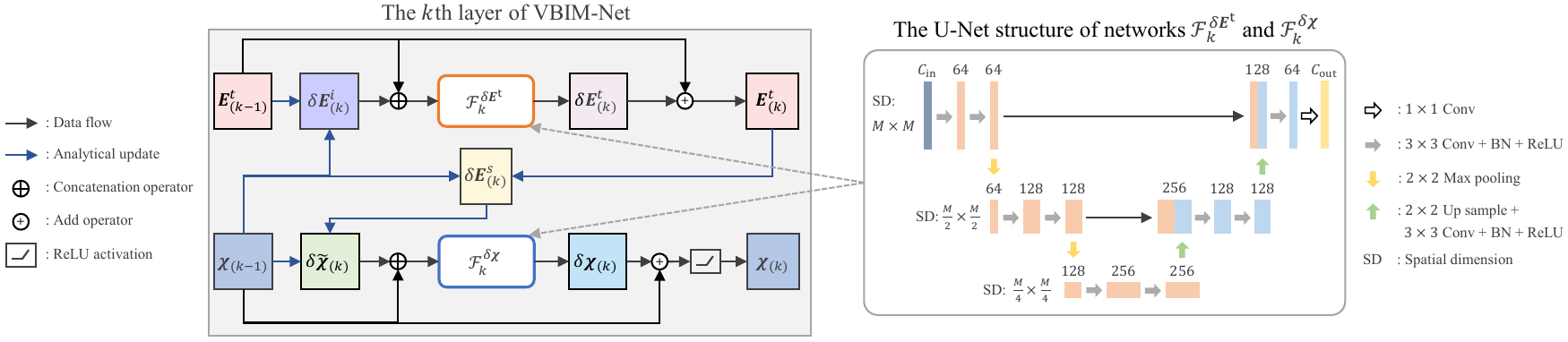}
    \captionsetup{font=footnotesize}
    \caption{The structure of each layer of VBIM-Net, in which the analytical update steps are indicated by blue arrows. 
    The networks $\mathcal{F}^{\delta\mathbf{E}^t}_{k}$ and $\mathcal{F}^{\delta\boldsymbol{\chi}}_{k}$ adopt U-Net with the same structure, as shown on the right side of the figure. 
    }
    \label{fig:VBIM_layer_and_UNet}
\end{figure*}

As depicted in Fig. \ref{fig:VBIM_Net_and_loss}, VBIM-Net includes $K$ layers of subnetworks to simulate the iterative process of VBIM. 
Assume that the initial value of contrast $\boldsymbol{\chi}_{(0)}$ is calculated by BPS, 
and the initial value of total field $\mathbf{E}^{t}_{(0)}$ is set according to the first-order Born approximation \cite{slaney1984Born}, i.e., $\mathbf{E}^{t}_{(0)} = \mathbf{E}^{i}$.  
Here, we concatenate the incident field $\mathbf{E}^{i}$ of $N_i$ incidences to obtain a real-value tensor with $2N_i$ channels as the input to VBIM-Net, 
and the network outputs the estimated total field for $N_i$ incidences at each layer. 
This allows the network to fully couple the data from multiple incidences to reconstruct the contrast of unknown scatterers. 
For contrast $\boldsymbol{\chi}$, we represent it with images with dimension $C\times M\times M$. 
For lossless scatters, the channel number $C$ is 1, while for lossy scatters, $C = 2$.     

The overall architecture of VBIM-Net refers to PhiSRL in \cite{shan2023PhiSRL}. 
To solve the matrix equation $\mathbf{Ax}=\mathbf{b}$, PhiSRL stacks multiple layers of neural networks with the same structure and residual connections to simulate the fixed-point iteration update process. 
In the $k$-th layer, the neural network computes
\begin{equation}
    \mathbf{x}_{(k)}=\mathbf{x}_{(k-1)}+\mathcal{F}_{k}(\delta_{(k-1)},\Theta_{k})=\mathbf{x}_{(k-1)}+\delta\mathbf{x}_{(k)}, 
\end{equation}
where $\delta_{(k-1)} = \mathbf{b} - \mathbf{A}\mathbf{x}_{(k-1)}$ is the equation residual from the $(k-1)$-th layer, 
$\mathcal{F}_{k}$ is the neural network in the $k$-th layer, 
and $\delta\mathbf{x}_{(k)} = \mathcal{F}_{k}(\delta_{(k-1)},\Theta_{k})$ is the predicted variation for $\mathbf{x}$. 

Each layer of VBIM-Net consists of a $\mathbf{E}^t$-update step and a $\boldsymbol{\chi}$-update step. 
In the $\mathbf{E}^t$-update step, instead of using the full-wave forward calculation in Eq. (\ref{eq:Et_update_BIM}), 
we adopt the form of one-step PhiSRL. 
Specifically, in the $k$th layer of the VBIM-Net, the $\mathbf{E}^t$-update step first calculates the incident field residual 
\begin{equation}
    \delta{\mathbf{E}^i_{(k)}} = \mathbf{E}^i - \left(\mathbf{I} - \mathbf{G}_{D} \operatorname{diag}\left({\boldsymbol{\chi}}_{(k-1)}\right)\right) {\mathbf{E}}^{t}_{(k-1)}, 
\end{equation}
and uses $\delta\mathbf{E}^i_{(k)} \oplus \mathbf{E}^{t}_{(k-1)}$ as the input of a NN, where $\oplus$ is the concatenation operation. 
The network output is the variation of the total field 
\begin{equation}
    \delta{\mathbf{E}^t_{(k)}} = \mathcal{F}^{\delta\mathbf{E}^t}_{k}\left(\delta\mathbf{E}^i_{(k)} \oplus \mathbf{E}^{t}_{(k-1)}; \boldsymbol{\Theta}^{\delta\mathbf{E}^t}_{k}\right), 
\end{equation}
where $\mathcal{F}^{\delta\mathbf{E}^t}_{k}$ denotes the NN for the calculation of total field variation and $\boldsymbol{\Theta}^{\delta\mathbf{E}^t}_{k}$ represents the corresponding network parameters.
Thus, the updated total field is
\begin{equation}
    \mathbf{E}^{t}_{(k)} = \mathbf{E}^{t}_{(k-1)} + \delta{\mathbf{E}^t_{(k)}}. 
\end{equation}
According to $\mathbf{E}^t_{(k)}$ and $\boldsymbol{\chi}_{(k-1)}$, we can obtain the scattered field residual $\delta\mathbf{E}^s_{(k)}$ for the $k$th iteration, 
\begin{equation}
    \delta{\mathbf{E}^s_{(k)}} = \mathbf{E}^s - \mathbf{G}_{S} \operatorname{diag}\left(\boldsymbol{\chi}_{(k-1)}\right) \mathbf{E}^{t}_{(k)}. 
\end{equation}

In the $\boldsymbol{\chi}$-update step, we also draw on the idea of PhiSRL, but based on the VBIM algorithm, we convert the residual $\delta{\mathbf{E}^s_{(k)}}$, which originally belongs to the field domain, into the contrast domain. 
Specifically, we first estimate the variation $\delta\boldsymbol{\chi}$ of the contrast according to Eq. (\ref{eq:delta_Es}). 
By stacking the data of $N_i$ incidences, we can transform Eq. (\ref{eq:delta_Es}) into the following linear equation, 
\begin{equation}
    \mathbf{y}_{(k)} = \mathbf{A}_{(k)} \delta\boldsymbol{\chi}, 
\end{equation}
where
\begin{equation}
    \mathbf{A}_{(k)} = \begin{bmatrix}
        \mathbf{G}_{S} \operatorname{diag}\left(\mathbf{E}^t_{(k), 1}\right) \\
        \vdots \\
        \mathbf{G}_{S} \operatorname{diag}\left(\mathbf{E}^t_{(k), N_i}\right)
    \end{bmatrix}, 
    \mathbf{y}_{(k)} = \operatorname{vec}\left(\delta{\mathbf{E}^s_{(k)}}\right), 
\end{equation}
and $\operatorname{vec}(\cdot)$ is the vectorization operator for a matrix.
The LS estimation of $\delta\boldsymbol{\chi}$ is 
\begin{equation}
    \delta\boldsymbol{\chi}^{\mathrm{LS}}_{(k)} = \left(\mathbf{A}_{(k)}^H \mathbf{A}_{(k)}\right)^{-1} \mathbf{A}_{(k)}^H \mathbf{y}_{(k)}, 
\end{equation}
which contains $M^2\times M^2$-dimensional matrix inversion with high computational complexity of $O(M^6)$. 
It is difficult to extend to high reconstruction resolutions.
Therefore, instead of using the LS estimation directly, we incorporate the idea of BPS and employ the following matched filter to estimate the approximate contrast variation, 
\begin{equation}
    \delta\tilde{\boldsymbol{\chi}}_{(k)} = \gamma^{\delta\tilde{\boldsymbol{\chi}}}_{(k)} \cdot \mathbf{A}_{(k)}^H \mathbf{y}_{(k)}, 
\end{equation}
where the coefficient $\gamma^{\delta\tilde{\boldsymbol{\chi}}}_{(k)}$ is obtained through LS estimation, 
\begin{equation}
    \gamma^{\delta\tilde{\boldsymbol{\chi}}}_{(k)} = \dfrac{\mathbf{y}_{(k)}^{T}\left(\mathbf{A}_{(k)} \mathbf{A}_{(k)}^H \mathbf{y}_{(k)}\right)^*}{\left\|\mathbf{A}_{(k)} \mathbf{A}_{(k)}^H \mathbf{y}_{(k)}\right\|^2}. 
\end{equation}
Although $\delta\tilde{\boldsymbol{\chi}}_{(k)}$ is a suboptimal estimate, it is computationally efficient and already contains the information from $\delta\mathbf{E}^s_{(k)}$. 
Next, we reshape $\delta\tilde{\boldsymbol{\chi}}_{(k)}$ into the dimension of $C\times M\times M$, 
and use $\delta\tilde{\boldsymbol{\chi}}_{(k)} \oplus \boldsymbol{\chi}_{(k-1)}$ as the input of a NN to get the refined variation $\delta\boldsymbol{\chi}_{(k)}$ of contrast, 
\begin{equation}
    \delta{\boldsymbol{\chi}_{(k)}} = \mathcal{F}^{\delta\boldsymbol{\chi}}_{k}\left(\delta\tilde{\boldsymbol{\chi}}_{(k)} \oplus \boldsymbol{\chi}_{(k-1)}; \boldsymbol{\Theta}^{\delta\boldsymbol{\chi}}_{k}\right), 
\end{equation}
where $\mathcal{F}^{\delta\boldsymbol{\chi}}_{k}$ denotes the NN for the calculation of contrast variation and $\boldsymbol{\Theta}^{\delta\boldsymbol{\chi}}_{k}$ represents the corresponding network parameters. 
Since the real and imaginary parts of the contrast are non-negative numbers, we use the rectified linear unit (ReLU) function to constrain $\boldsymbol{\chi}_{(k)}$. 
Therefore, the updated contrast is
\begin{equation}
    \boldsymbol{\chi}_{(k)} = \left(\boldsymbol{\chi}_{(k-1)} + \delta\boldsymbol{\chi}_{(k)}\right)^+, 
\end{equation}
where $(\cdot)^+$ denotes the ReLU function. 
Given the convergence advantage of VBIM over BIM, our proposed approximate contrast variation more efficiently conveys the information needed for contrast updating than the scattered field residuals in NeuralBIM, while naturally avoiding the $N_r\times N_i = M\times M$ constraint. 
% This design naturally avoids the requirement for $N_r\times N_i = M\times M$ in NeuralBIM and aids the network to predict the contrast variation. 

The structure of each layer of VBIM-Net is shown in Fig. \ref{fig:VBIM_layer_and_UNet}. 
We adopt U-Net \cite{ronneberger2015UNet} to build NNs $\mathcal{F}^{\delta\mathbf{E}^t}_{k}$ and $\mathcal{F}^{\delta\boldsymbol{\chi}}_{k}$ for predicting the variation $\delta{\mathbf{E}^t_{(k)}}$ and $\delta{\boldsymbol{\chi}_{(k)}}$, 
which efficiently integrates multi-level feature information and leverages the similarity between input and output, and has been widely applied to solve nonlinear ISPs \cite{wei2018DLS}, \cite{wei2019ICLM}, \cite{wang2023MSDLS}, \cite{wang2023MDWS}.
The spatial dimension and channel number of each layer of U-Net are marked in Fig. \ref{fig:VBIM_layer_and_UNet}, where the number of input channel and output channel are denoted by $C_{\mathrm{in}}$ and $C_{\mathrm{out}}$, respectively. 
For the total field update network $\mathcal{F}^{\delta\mathbf{E}^t}_{k}$, $C_{\mathrm{in}} = 4N_i, C_{\mathrm{out}} = 2N_i$. 
For the contrast update network $\mathcal{F}^{\delta\boldsymbol{\chi}}_{k}$, $C_{\mathrm{in}} = 2, C_{\mathrm{out}} = 1$ for lossless scatterers, and $C_{\mathrm{in}} = 4, C_{\mathrm{out}} = 2$ for lossy scatterers.
The CNNs in the U-Net consist of $3\times 3$ convolutions, batch normalization (BN), and ReLU activation functions,  
and the final output layers use $1\times 1$ convolution to project the feature map to the output space. 
Algorithm \ref{algo:VBIM_net} summarizes the data flow of VBIM-Net. 
In this paper, VBIM-Net is assumed to consist of seven layers of subnetworks ($K=7$), which share the same structure but have independent parameters. 

\begin{algorithm}[!t]
\caption{The data flow in the VBIM-Net}\label{algo:VBIM_net}
\begin{algorithmic}[1]
\setlength{\itemsep}{0.5em} 
\Require Scattered field $\mathbf{E}^s$, incident field $\mathbf{E}^i$, Green's function matrix $\mathbf{G}_D, \mathbf{G}_S$, the number of network layers $K$.
\Ensure The predicted contrast $\hat{\boldsymbol{\chi}}$ and total field $\hat{\mathbf{E}}^t$. 
\State \textbf{Initialization}: Use the BPS algorithm to calculate the initial value $\boldsymbol{\chi}_{(0)}$ according to Eq.(\ref{eq:BPS_ICC})-(\ref{eq:BPS_x}). $\mathbf{E}^t_{(0)} = \mathbf{E}^i$.
\For{$k = 1,\cdots,K$}
    \State $\delta{\mathbf{E}^i_{(k)}} = \mathbf{E}^i - \left(\mathbf{I} - \mathbf{G}_{D} \operatorname{diag}\left({\boldsymbol{\chi}}_{(k-1)}\right)\right) {\mathbf{E}}^{t}_{(k-1)}$
    \State $\delta{\mathbf{E}^t_{(k)}} = \mathcal{F}^{\delta\mathbf{E}^t}_{k}\left(\delta\mathbf{E}^i_{(k)} \oplus \mathbf{E}^{t}_{(k-1)}; \boldsymbol{\Theta}^{\delta\mathbf{E}^t}_{k}\right)$
    \State $\mathbf{E}^t$-update: $\mathbf{E}^{t}_{(k)} = \mathbf{E}^{t}_{(k-1)} + \delta{\mathbf{E}^t_{(k)}}$
    \State $\delta{\mathbf{E}^s_{(k)}} = \mathbf{E}^s - \mathbf{G}_{S} \operatorname{diag}\left(\boldsymbol{\chi}_{(k-1)}\right) \mathbf{E}^{t}_{(k)}$
    \State $\mathbf{A}_{(k)} = \begin{bmatrix}
        \mathbf{G}_{S} \operatorname{diag}\left(\mathbf{E}^t_{(k), 1}\right) \\
        \vdots \\
        \mathbf{G}_{S} \operatorname{diag}\left(\mathbf{E}^t_{(k), N_i}\right)
    \end{bmatrix}, \mathbf{y}_{(k)} = \operatorname{vec}\left(\delta{\mathbf{E}^s_{(k)}}\right)$
    \State $\gamma^{\delta\tilde{\boldsymbol{\chi}}}_{(k)} = \dfrac{\mathbf{y}_{(k)}^{T}\left(\mathbf{A}_{(k)} \mathbf{A}_{(k)}^H \mathbf{y}_{(k)}\right)^*}{\left\|\mathbf{A}_{(k)} \mathbf{A}_{(k)}^H \mathbf{y}_{(k)}\right\|^2}$
    \State $\delta\tilde{\boldsymbol{\chi}}_{(k)} = \gamma^{\delta\tilde{\boldsymbol{\chi}}}_{(k)} \cdot \mathbf{A}_{(k)}^H \mathbf{y}_{(k)}$
    \State $\delta{\boldsymbol{\chi}_{(k)}} = \mathcal{F}^{\delta\boldsymbol{\chi}}_{k}\left(\delta\tilde{\boldsymbol{\chi}}_{(k)} \oplus \boldsymbol{\chi}_{(k-1)}; \boldsymbol{\Theta}^{\delta\boldsymbol{\chi}}_{k}\right)$
    \State $\boldsymbol{\chi}$-update : $\boldsymbol{\chi}_{(k)} = \left(\boldsymbol{\chi}_{(k-1)} + \delta\boldsymbol{\chi}_{(k)}\right)^+$
\EndFor
\State $\hat{\boldsymbol{\chi}} = \boldsymbol{\chi}_{(K)}$, $\hat{\mathbf{E}}^t = \mathbf{E}^t_{(K)}$
\end{algorithmic}
\end{algorithm}

\subsection{Loss Function} \label{subsec:loss}
In order to make each layer output of VBIM-Net corresponds to the actual physical quantity, 
we designed a layer-wise constraint in the loss function, which explicitly constrains the output contrast and total field of each layer, decaying the importance of each layer's output from back to front.
The overall loss function is formulated as
\begin{equation} \label{eq:loss}
    \mathcal{L}_{\boldsymbol{\chi},\mathbf{E}^t} = \sum_{k=1}^{K}{w_k \cdot \mathcal{L}^{\boldsymbol{\chi},\mathbf{E}^t}_k},
\end{equation}
where $\mathcal{L}^{\boldsymbol{\chi},\mathbf{E}^t}_k$ is the loss on the predicted contrast and total field for the $k$th subnetwork, 
and $w_k$ represents the weight of the $k$th layer. 
For the layer weight $w_k$, we adopt the form of exponential, i.e., 
\begin{equation} \label{eq: w_k}
    w_k = c^{K-1-k}, \quad k = 1, \cdots, K, 
\end{equation}
with parameter $c \in [0, 1)$ as a decay factor, which is set to 0.8 by default.  
In this way, the outputs of each subnetwork in VBIM-Net correspond to actual physical quantities, 
while also providing reasonable residual $\delta\mathbf{E}^i$ and $\delta\mathbf{E}^s$ for the next subnetwork. 
In general, the layer-wise constraint can be interpreted as a regularization technique based on soft physical constraints, which assists the data flow of VBIM-Net to consistent with the paradigm of iterative algorithms. In the following section, we will verify the effectiveness and necessity of the layer-wise constraint in training deep VBIM-Net models. 

In Eq. (\ref{eq:loss}), the loss for the $k$th layer of VBIM-Net is
\begin{equation} \label{eq:layer_loss}
    \mathcal{L}^{\boldsymbol{\chi},\mathbf{E}^t}_k = \mathcal{L}^{\boldsymbol{\chi}}_k + \alpha \cdot \mathcal{L}^{\boldsymbol{\mathbf{E}^t}}_k, 
\end{equation}
where the coefficient $\alpha$ is used to balance the importance of $\boldsymbol{\chi}$ and $\mathbf{E}^t$ and it is set to $0.5$ in this paper.
$\mathcal{L}^{\boldsymbol{\mathbf{E}^t}}_k$ is the mean square error (MSE) of $\mathbf{E}^t_{(k)}$ output by the $k$th layer, 
\begin{equation}
    \mathcal{L}^{\boldsymbol{\mathbf{E}^t}}_k = \frac{1}{N_{\mathbf{E}^t}} \left\|\mathbf{E}^t_{(k)} - \mathbf{E}^{t,*}\right\|_F^2,
\end{equation}
where $N_{\mathbf{E}^t}$ denotes the element number of $\mathbf{E}^t$, $\mathbf{E}^{t,*}$ denotes the ground truth of the total field, and $\|\cdot\|_F$ is the Frobenius norm.
$\mathcal{L}^{\boldsymbol{\chi}}_k$ is the MSE loss of $\boldsymbol{\chi}_{(k)}$ with total variation (TV) regularization, 
\begin{equation}
    \mathcal{L}^{\boldsymbol{\chi}}_k = \frac{1}{N_{\boldsymbol{\chi}}} \left\|\boldsymbol{\chi}_{(k)} - \boldsymbol{\chi}^{*}\right\|_{F}^2 + \beta \cdot \frac{1}{N_{\boldsymbol{\chi}}}\left\|\nabla\boldsymbol{\chi}_{(k)}\right\|_1, 
\end{equation}
where TV regularization is introduced to improve the spatial continuity of predicted contrast images and alleviate the ill-posedness of ISPs. 
$N_{\boldsymbol{\chi}}$ denotes the element number of $\boldsymbol{\chi}$, $\boldsymbol{\chi}^{*}$ denotes the ground truth of the contrast, $\|\cdot\|_1$ denotes the L1-norm, and $\beta$ is fixed as 0.0001.
The calculation process of the loss function is depicted in Fig. \ref{fig:VBIM_Net_and_loss}. 

To enhance the model's robustness against noise, we randomly add Gaussian noise with a certain range of SNR to the scattered field measurements obtained from forward simulation, imitating actual measurement data in the training process. 
At this time, in order to enable the model to distinguish training data with different noise levels, we designed a loss weighting strategy proportional to the SNR of training samples. 
According to the definition of the noise level, it can be expressed as
\begin{equation} \label{eq:nl_weight}
\begin{aligned}
    \mathcal{L}_{\mathrm{opt}} &= \gamma \cdot \mathrm{SNR} \cdot \mathcal{L}_{\boldsymbol{\chi},\mathbf{E}^t}\\
    &= \frac{\gamma}{noise\_level^2} \cdot \mathcal{L}_{\boldsymbol{\chi},\mathbf{E}^t}
\end{aligned}
\end{equation}
where $\mathcal{L}_{\mathrm{opt}}$ is the final loss for optimizing the model parameters and the coefficient $\gamma$ is empirically set to 0.04. 
This training scheme allows the model to adapt to variational noise levels and focus more on the reconstruction results at low noise levels.

\subsection{Complexity Analysis} \label{subsec:complexity}
The computational complexity of VBIM-Net mainly consists of five parts: 
\begin{itemize}
    \item Preparing the initial value $\boldsymbol{\chi}_{(0)}$ of contrast through the BPS algorithm.
    \item Calculating the incident field residual $\delta\mathbf{E}^i$. 
    \item Calculating the scattered field residual $\delta\mathbf{E}^s$.
    \item Calculating the approximate variation $\delta\tilde{\boldsymbol{\chi}}$ of contrast.
    \item Updating the $\boldsymbol{\chi}$ and $\mathbf{E}^t$ via the U-Nets.
\end{itemize}

The main computational complexity of the BPS algorithm comes from calculating $\mathbf{G}_D \mathbf{J}^{b}$, 
which is $O(N_i M^2 \log{M^2})$ \cite{wei2018DLS} if the fast Fourier transform (FFT) is applied in the matrix–vector multiplication.

The data flow of each VBIM-Net layer is divided into the $\mathbf{E}^t$-update branch and the $\boldsymbol{\chi}$-update branch. 
In the $\mathbf{E}^t$-update branch, $\delta\mathbf{E}^i$ needs to be calculated first, and its complexity mainly comes from the $\mathbf{G}_D \operatorname{diag}\left(\boldsymbol{\chi}\right)\mathbf{E}^t$ term, which is $O(N_i M^2 \log{M^2})$ if the FFT is applied in the matrix–vector multiplication.
The computational complexity of $\delta\mathbf{E}^s$ calculation mainly comes from the matrix multiplication contained in the $\mathbf{G}_S \operatorname{diag}\left(\boldsymbol{\chi}\right)\mathbf{E}^t$ term, which is $O(N_r N_i M^2)$. 
In the $\boldsymbol{\chi}$-update branch, the approximate contrast variation $\delta\tilde{\boldsymbol{\chi}}$ needs to be calculated first, 
and its complexity mainly comes from the preparation of matrix $\mathbf{A}$, the $\mathbf{A}^H \mathbf{y}$ calculation, and the $\mathbf{A}(\mathbf{A}^H \mathbf{y})$ calculation, all of which are  $O(N_r N_i M^2)$, and we store the value of $\mathbf{A}^H \mathbf{y}$ to avoid duplicate computations. 

The forward calculation of U-Net mainly consists of several basic operations, including convolution, activation function, and max pooling, where the computational complexity is dominated by convolutions. 
In the convolution operation, if the feature map size, the convolution kernel size, the number of input channels and output channels are denoted by $H\times W$, $K_h\times K_w$, $C_i$, and $C_o$, respectively, then the complexity is $O(C_i C_o H W K_h K_w)$.

\begin{figure}[!t]
    \centering
    \includegraphics[width=\linewidth]{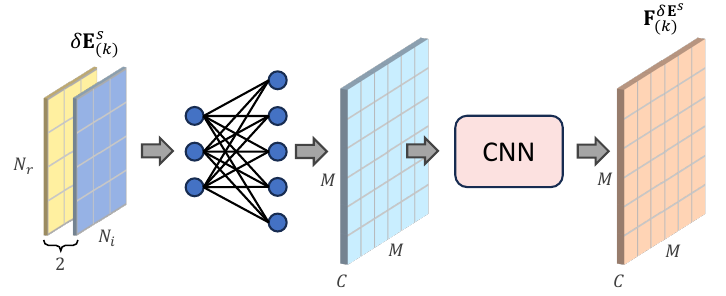}
    \captionsetup{font=footnotesize}
    \caption{Module executing the mapping $\delta\mathbf{E}^s \rightarrow \mathbf{F}^{\delta\mathbf{E}^s}$ for dimension transformation in each layer of the modified NeuralBIM.}
    \label{fig:NeuralBIM_proj}
\end{figure}

\begin{figure}[t]
    \centering
    \includegraphics[width=\linewidth]{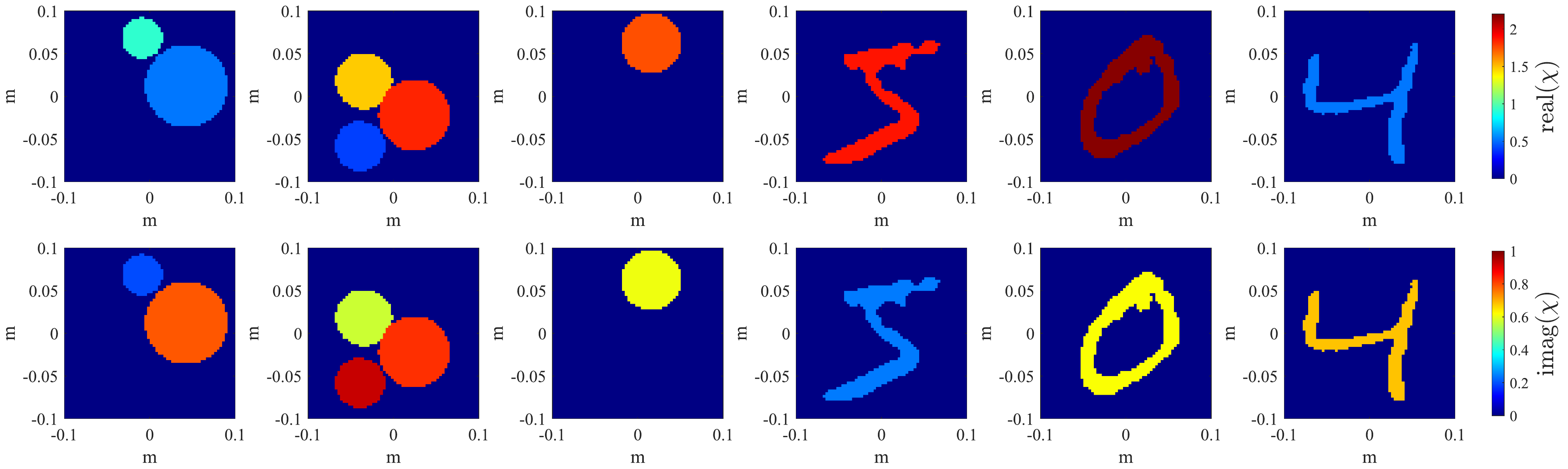}
    \captionsetup{font=footnotesize}
    \caption{Examples of scatterers in the synthetic dataset for lossy scatterers. 
    The first row also represents the examples from the lossless dataset.}
    \label{fig:dataset_example}
\end{figure}

\section{Numerical Results} \label{sec:numerical_results}
In this section, we demonstrate the effectiveness of VBIM-Net on synthetic and experimental data,
evaluating the model's reconstruction quality, generalization ability, and robustness. 
Based on this, we designed a series of experiments to validate the physical insight and experimental efficacy of the new designs in proposed VBIM-Net. 

\subsection{Benchmark Algorithms and Performance Metrics} \label{subsec:benchmark_metric}
We select BIM, VBIM, SOM \cite{chen2009SOM} and NeuralBIM \cite{shan2022NeuralBIM} as benchmark algorithms. 
The BIM and VBIM algorithm are introduced in Section \ref{subsec:BIM}, and the L2 regularization parameter $\lambda$ is empirically set to $5\times 10^{-4}$ here. 

SOM is a classic source-type iterative method that decomposes the contrast current into a deterministic part and an ambiguous part through SVD, 
and uses a two-step CG algorithm to alternately optimize the contrast and ambiguous current component, 
which has good inversion performance in most scenarios \cite{chen2018computational}, \cite{chen2009SOM}, \cite{liu2022SOMNet}. 

NeuralBIM is proposed in \cite{shan2022NeuralBIM}, which is a representative field-type DL scheme that unrolls BIM and designs the update networks for contrast $\boldsymbol{\chi}$ and total field $\mathbf{E}^t$ based on PhiSRL \cite{shan2023PhiSRL}.
Specifically, it uses the residuals of Eq. (\ref{eq:discrete_state_eq}) and (\ref{eq:discrete_data_eq}) as inputs of two ResNets respectively, simulating the fixed-point iterative method to solve $\boldsymbol{\chi}$ and $\mathbf{E}^t$. 
However, the vanilla NeuralBIM simply uses $\delta\mathbf{E}^s \oplus \boldsymbol{\chi}$ as the input of $\boldsymbol{\chi}$ update network, 
which only suitable for cases where the scattered field dimension matches the reconstruction grid resolution \cite{shan2022NeuralBIM}. 
In contrast, the proposed VBIM-Net transfers the information of residual $\delta\mathbf{E}^s$ through the approximate contrast variation $\delta\tilde{\boldsymbol{\chi}}$, 
enabling it to handle more flexible measurement data dimensions and grid resolutions inherently. 
To facilitate comparison with VBIM-Net, this paper made some modifications to NeuralBIM. 
We add a dimension transformation module as shown in Fig. \ref{fig:NeuralBIM_proj} to each layer of NeuralBIM, 
which maps $\delta\mathbf{E}^s$ with dimension $2\times N_r\times N_i$ to a feature map $\mathbf{F}^{\delta\mathbf{E}^s}$ with dimension $C\times M\times M$ through a cascaded fully connected layer and CNN. 
We use $\mathbf{F}^{\delta\mathbf{E}^s} \oplus \boldsymbol{\chi}$ as the input of $\boldsymbol{\chi}$-update subnetworks in the modified NeuralBIM. 
For the sake of fairness, we let the modified NeuralBIM have the same number of update layers as VBIM-Net and use the same dataset for training. 
The modified NeuralBIM adopts the supervised learning scheme in \cite{shan2022NeuralBIM}, which only supervises the final reconstruction results of contrast and electric field. Compared with the loss function of VBIM-Net, this original scheme lacks physical constraints on the parameters of the intermediate layers, resulting in the data flow of NeuralBIM not conforming well to the iterative regime, thus degrading the model’s physical interpretability and generalization ability. 

In order to evaluate the reconstruction quality, we take the normalized mean squared error (NMSE) and the structural similarity index measure (SSIM) as performance metrics.
The definition of the NMSE is
\begin{equation}
    \operatorname{NMSE}(\boldsymbol{\chi}, \boldsymbol{\chi}^*) = \frac{\left\|\boldsymbol{\chi}-\boldsymbol{\chi}^{*}\right\|_{F}^2}{\left\|\boldsymbol{\chi}^{*}\right\|_{F}^2} , 
\end{equation}
and the definition of SSIM \cite{wang2004SSIM} is
\begin{equation}
    \operatorname{SSIM}(\boldsymbol{\chi}, \boldsymbol{\chi}^*) = \frac{(2\mu_{\boldsymbol{\chi}}\mu_{\boldsymbol{\chi}^*}+C_1)(2\sigma_{\boldsymbol{\chi}\boldsymbol{\chi}^*}+C_2)}{(\mu_{\boldsymbol{\chi}}^2+\mu_{\boldsymbol{\chi}^*}^2+C_1)(\sigma_{\boldsymbol{\chi}}^2+\sigma_{\boldsymbol{\chi}^*}^2+C_2)}, 
\end{equation}
where $\mu_{\boldsymbol{\chi}}$ and $\sigma_{\boldsymbol{\chi}}$ are the mean and standard deviation of the elements in $\boldsymbol{\chi}$, respectively. 
$\sigma_{\boldsymbol{\chi}\boldsymbol{\chi}^*}$ is the covariance between $\boldsymbol{\chi}$ and $\boldsymbol{\chi}^*$. 
$C_1 = (K_1 L_{\boldsymbol{\chi}})^2$, $C_2 = (K_2 L_{\boldsymbol{\chi}})^2$ are parameters used to ensure the stability of the division, where $L_{\boldsymbol{\chi}}$ is the dynamic range of $\boldsymbol{\chi}$, and $k_1=0.01, k_2=0.03$ by default.

\begin{figure*}[!htbp]
    \centering
    \begin{subfigure}[t]{0.99\linewidth}
        \includegraphics[width=\linewidth]{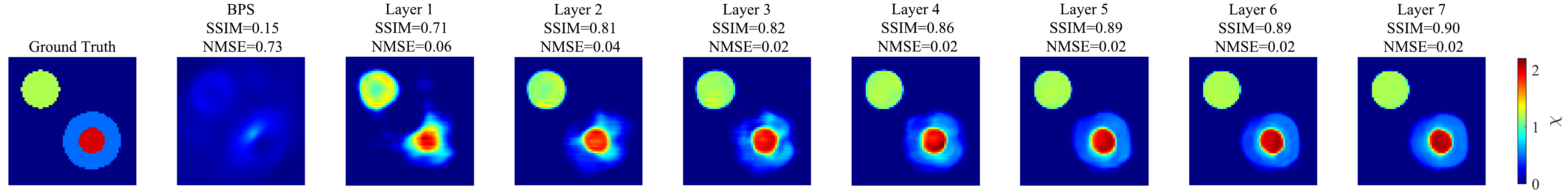}
        \captionsetup{font=footnotesize}
        \caption{Cylinder dataset example.}
        \vspace{2pt}
        \label{fig:recon_layers_cylinder}
    \end{subfigure}
    \begin{subfigure}[t]{0.99\linewidth}
        \includegraphics[width=\linewidth]{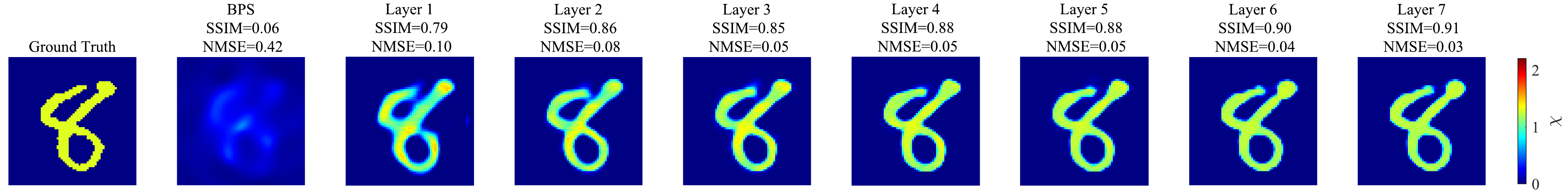}
        \captionsetup{font=footnotesize}
        \caption{MNIST dataset example.}
        \label{fig:recon_layers_MNIST}
    \end{subfigure}
    \captionsetup{font=footnotesize}
    \caption{Updated contrast reconstructions of dataset examples of VBIM-Net. 
    From left to right, the ground truth of contrast, the initial value obtained by BPS, 
    and the reconstruction result of each layer are shown. The constant $c = 0.8$. 
    }
    \label{fig:recon_layers}
\end{figure*}

\subsection{System Configurations and Training Details} \label{subsec:implementation_details}

The system configurations of the synthetic data are as follows: 
$N_i$ line sources and $N_r$ line receivers are equally placed on a circle with a radius of 1.67 m, 
where $N_i = 16$, $N_r = 32$ by default, and the other measurement dimensions will be discussed in Section \ref{subsec:synthetic_cross}. 
The frequency of the incident wave is 3 GHz. 
$D$ is the DOI of size $0.2 \mathrm{m} \times 0.2 \mathrm{m}$ with a free space background. 
In the forward problem, $D$ is discretized into a grid of $100\times 100$, and the MoM is used to simulate the scattered field $\mathbf{E}^s$. 
In the inverse problem, we need to discretize $D$ at a lower resolution than the forward problem to avoid the inverse crime\cite{chen2018computational}, and $64\times 64$ is chosen in this paper. 
The simulated $\mathbf{E}^s$ is noise-free, so we add different levels of Gaussian noise to $\mathbf{E}^s$ to simulate the actual measurement data. 
The noise level is defined as $(\|\mathbf{N}\|_F / \|\mathbf{E}^s\|_F)$, where $\mathbf{N}$ is the noise matrix. 
During training, we randomly add 5\%-35\% noise to $\mathbf{E}^s$ and employ the weighted loss in Eq. (\ref{eq:nl_weight}). 
By default, we assume that the scatterers are lossless, and we will discuss the inversion results for lossy scatterers separately in Section \ref{subsec:synthetic_cross}. 
The system and dataset configuration for the experimental data is slightly different, which will be described in Section \ref{subsec:experimental}.

We mix the MNIST \cite{lecun1998MNIST} and cylinder dataset at a ratio of 1:1 to obtain the synthetic dataset, 
in order to increase the richness of the data, making the network to learn inversion tasks for homogeneous and inhomogeneous scatterers with various shapes. 
For the cylinder dataset, each scene contains 1 to 3 cylinders with a 0.01m-0.05m radius. 
For lossless scatters, the contrast range is $[0.2,2.2]$. For lossy scatters, the range of the real part of the contrast is $[0.2,2.2]$, and the range of the imaginary part is $[0,1]$.
Fig. \ref{fig:dataset_example} shows several examples from the lossy dataset, with the first row in the figure also representing examples from the lossless dataset. 
Unless otherwise specified, in the subsequent experiments, we use the lossless dataset for training and testing by default.
We randomly generate 20,000 data samples, of which 80\% are for training and 20\% for testing. 

VBIM-Net and the modified NeuralBIM are implemented in PyTorch and computed on a Nvidia V100 GPU, and they share the same training configuration. 
Table \ref{tab:implementation_detail} summarizes the implementation details of both, 
and Table \ref{tab:params_train_infer_time} lists the number of parameters, total training time and single-sample inference time for both. 
Note that since the modified NeuralBIM adds a trainable dimension transformation module to implement the mapping $\delta\mathbf{E}^s \rightarrow \mathbf{F}^{\delta\mathbf{E}^s}$, 
the number of trainable parameters of the modified NeuralBIM is larger than that of VBIM-Net, which also leads to a longer training time. 
However, during the inference stage, the $\boldsymbol{\chi}$-update network of VBIM-Net contains an analytical computation of the approximate contrast variation $\delta\tilde{\boldsymbol{\chi}}$, resulting in a longer inference time.

\begin{table}[htbp]
\centering
\captionsetup{font=footnotesize}
\caption{Implementation Details of VBIM-Net and the Modified NeuralBIM} \label{tab:implementation_detail}
\renewcommand{\arraystretch}{1.1}
\begin{tabular}{|l|l|}
\hline
The number of layers ($K$)        & 7                   \\ \hline
Training epochs                   & 80                  \\ \hline
Optimizer                         & Adam                \\ \hline
Initial learning rate             & $1\times 10^{-4}$   \\ \hline
Learning rate schedular           & $\times 0.8$ every 10 epochs     \\ \hline
\end{tabular}
\end{table}

\begin{table}[htbp]
\centering
\captionsetup{font=footnotesize}
\caption{The number of parameters, total training time and single-sample inference time for modified NeuralBIM and VBIM-Net} \label{tab:params_train_infer_time}
\renewcommand{\arraystretch}{1.5}
\begin{tabular}{|c|c|c|c|}
\hline
{Method}  & {Parameters}   & {\parbox{2.1cm}{Total training time}}     & {\parbox{2cm}{Single-sample\\ inference time}}         \\ \hline
NeuralBIM    &  48.62 Million  & 16h       & 0.0178s      \\ \hline
VBIM-Net     &  29.24 Million  & 10.58h    & 0.0273s      \\ \hline
\end{tabular}
\end{table}

\begin{figure*}[!t]
    \centering
    \includegraphics[width=\linewidth]{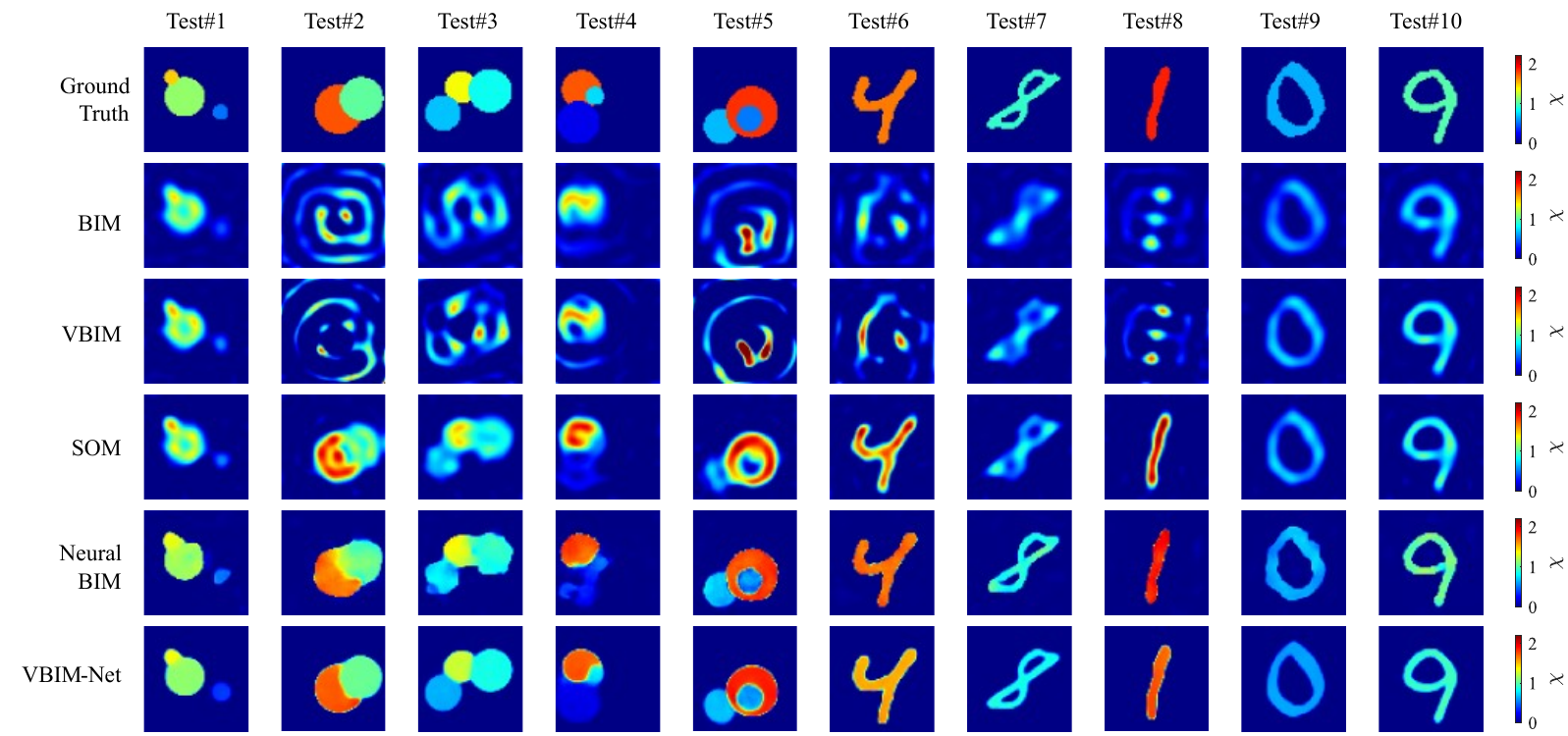}
    \captionsetup{font=footnotesize}
    \caption{Reconstruction results of within-dataset examples Test\#1-Test\#10 from the cylinder and MNIST test sets with 10\% Gaussian noise.}
    \label{fig:within_dataset_examples}
\end{figure*}

\begin{table*}[htbp]
\captionsetup{font=footnotesize}
\caption{Performance Metrics of Reconstruction Results for Test\#1-Test\#10} \label{tab:performance_within_dataset_example}
\centering
\renewcommand{\arraystretch}{1.1}
\begin{tabular}{|c|c|c|c|c|c|c|c|c|c|c|}
\hline
\multirow{2}{*}{Method} & \multicolumn{2}{c|}{Test\#1} & \multicolumn{2}{c|}{Test\#2} & \multicolumn{2}{c|}{Test\#3} & \multicolumn{2}{c|}{Test\#4} & \multicolumn{2}{c|}{Test\#5}\\
\cline{2-11}
                                 & SSIM & NMSE & SSIM & NMSE & SSIM & NMSE & SSIM & NMSE & SSIM & NMSE \\
\hline
\multicolumn{1}{|c|}{BIM}        & 0.60 & 0.13 & 0.11 & 0.74 & 0.29 & 0.44 & 0.52 & 0.25 & 0.19 & 0.74\\
\hline
\multicolumn{1}{|c|}{VBIM}        & 0.76 & 0.10 & 0.24 & 0.97 & 0.36 & 0.68 & 0.55 & 0.37 & 0.27 & 1.01\\
\hline
\multicolumn{1}{|c|}{SOM}        & 0.80 & 0.10 & 0.68 & 0.07 & 0.73 & 0.08 & 0.76 & 0.11 & 0.70 & 0.10\\
\hline
\multicolumn{1}{|c|}{NeuralBIM} & 0.81 & 0.09 & 0.79 & 0.06 & 0.78 & 0.09 & 0.74 & 0.11 & 0.88 & 0.03\\
\hline
\multicolumn{1}{|c|}{VBIM-Net} & \textbf{0.94} & \textbf{0.04} & \textbf{0.95} & \textbf{0.02} & \textbf{0.93} & \textbf{0.04} & \textbf{0.92} & \textbf{0.07} & \textbf{0.95} & \textbf{0.02} \\
\hline
\hline
\multirow{2}{*}{Method} & \multicolumn{2}{c|}{Test\#6} & \multicolumn{2}{c|}{Test\#7} & \multicolumn{2}{c|}{Test\#8} & \multicolumn{2}{c|}{Test\#9} & \multicolumn{2}{c|}{Test\#10}\\
\cline{2-11}
                                 & SSIM & NMSE & SSIM & NMSE & SSIM & NMSE & SSIM & NMSE & SSIM & NMSE \\
\hline
\multicolumn{1}{|c|}{BIM}        & 0.21 & 0.84 & 0.52 & 0.37 & 0.34 & 0.57 & 0.58 & 0.19 & 0.52 & 0.29\\
\hline
\multicolumn{1}{|c|}{VBIM}        & 0.31 & 1.00 & 0.71 & 0.30 & 0.47 & 0.68 & 0.79 & 0.12 & 0.76 & 0.17\\
\hline
\multicolumn{1}{|c|}{SOM}        & 0.82 & 0.11 & 0.78 & 0.27 & 0.88 & 0.11 & 0.83 & 0.11 & 0.82 & 0.15\\
\hline
\multicolumn{1}{|c|}{NeuralBIM} & 0.91 & \textbf{0.03} & 0.83 & 0.13 & 0.87 & \textbf{0.06} & 0.84 & 0.10 & 0.84 & 0.10\\
\hline
\multicolumn{1}{|c|}{VBIM-Net} & \textbf{0.95} & {0.05} & \textbf{0.95} & \textbf{0.08} & \textbf{0.97} & \textbf{0.06} & \textbf{0.95} & \textbf{0.06} & \textbf{0.95} & \textbf{0.07} \\
\hline
\end{tabular}
\end{table*}

\begin{figure}[!t]
    \centering
    \vspace{-0.2cm}
    \includegraphics[width=\linewidth]{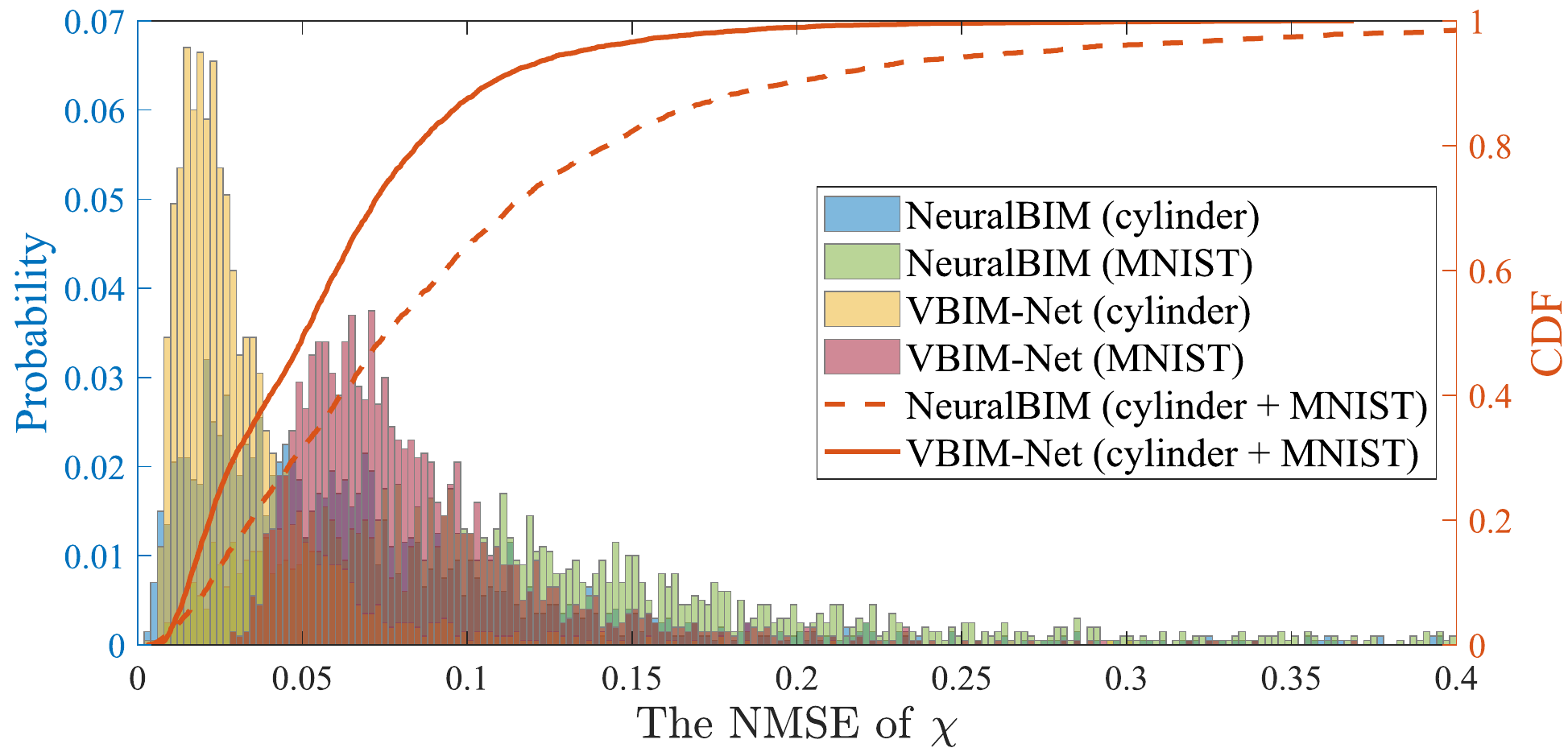}
    \captionsetup{font=footnotesize}
    \caption{The histograms and CDFs shows the NMSE distribution of reconstructed contrast images obtained by NeuralBIM and VBIM-Net 
    on test sets, where 10\% Gaussian noise is added.}
    \label{fig:test_nmse_2algo}
\end{figure}

\subsection{Synthetic Data Inversion: Within-Dataset Test} \label{subsec:synthetic_within}

In this subsection, we first visually display the output of each VBIM-Net layer, and then test the model's performance on the MNIST and cylinder datasets. 
Finally, we conduct a series of experiments to verify the efficacy of the subnetwork input configuration and the layer-wise constraint in VBIM-Net. 

\subsubsection{Updated Contrast Reconstructions of VBIM-Net}
Fig. \ref{fig:recon_layers} illustrates the updated contrast reconstructions of VBIM-Net for the cylinder and MNIST dataset examples. 
The initial value of contrast is obtained by BPS, and it is gradually improved by the update layers of VBIM-Net. 
Due to the layer-wise constraint in the loss function, the output $\boldsymbol{\chi}_{(k)}$ of each layer of VBIM-Net is a reasonable contrast image, which also provides reasonable residuals of incident field $\delta\mathbf{E}^i_{(k+1)}$ and scattered field $\delta\mathbf{E}^s_{(k+1)}$ for the next layer.
It guarantees the variables in VBIM-Net to have expected physical meanings, ensuring the reliability of the model. 

\subsubsection{Within-Dataset Samples Reconstruction}
Fig. \ref{fig:within_dataset_examples} shows several reconstruction results on the cylinder and MNIST test sets of these three algorithms when the noise level is 10\%, 
and the corresponding performance metrics are summarized in Table \ref{tab:performance_within_dataset_example}. 
Reconstruction results indicate that VBIM-Net outperforms all the comparison algorithms. 
Although both VBIM-Net and NeuralBIM are designed based on the field-type iterative method, the reconstruction quality of VBIM-Net is significantly better. 
This is mainly attributed to the analytical calculation of the approximate contrast variation $\delta\tilde{\boldsymbol{\chi}}$ that converts the scattered field residual $\delta\mathbf{E}^s$ into the contrast domain, making it more conducive to feature extraction by the subsequent CNN. 
Moreover, the U-Net used by VBIM-Net is more effective in extracting the multi-level features in contrast images than the ResNet in NeuralBIM, which also helps to improve the model performance.

Fig. \ref{fig:test_nmse_2algo} displays the histograms and cumulative distribution functions (CDFs) of the NMSE of reconstructed contrast images obtained by NeuralBIM and VBIM-Net. The noise level is 10\%. 
In Fig. \ref{fig:test_nmse_2algo}, the histograms are drawn for MNIST and cylinder scatterers separately to reflect the differences between the two datasets. 
In contrast, the CDF curves are drawn based on the entire test set to compare the overall performance of these two DL methods. 
Judging from the distribution of NMSE, VBIM-Net has more minor errors on test sets and a more concentrated error distribution, which means it performs better than NeuralBIM. 
Besides, it can be found that there is a significant difference in the performance of the two models on the MNIST and cylinder datasets. 
Therefore, mixing them in the training stage can enhance the data richness. 

\subsubsection{Validation of the Efficacy of the Subnetwork Input Configuration}
\begin{figure}[htbp]
    \centering
    \vspace{-0.2cm}
    \includegraphics[width=0.95\linewidth]{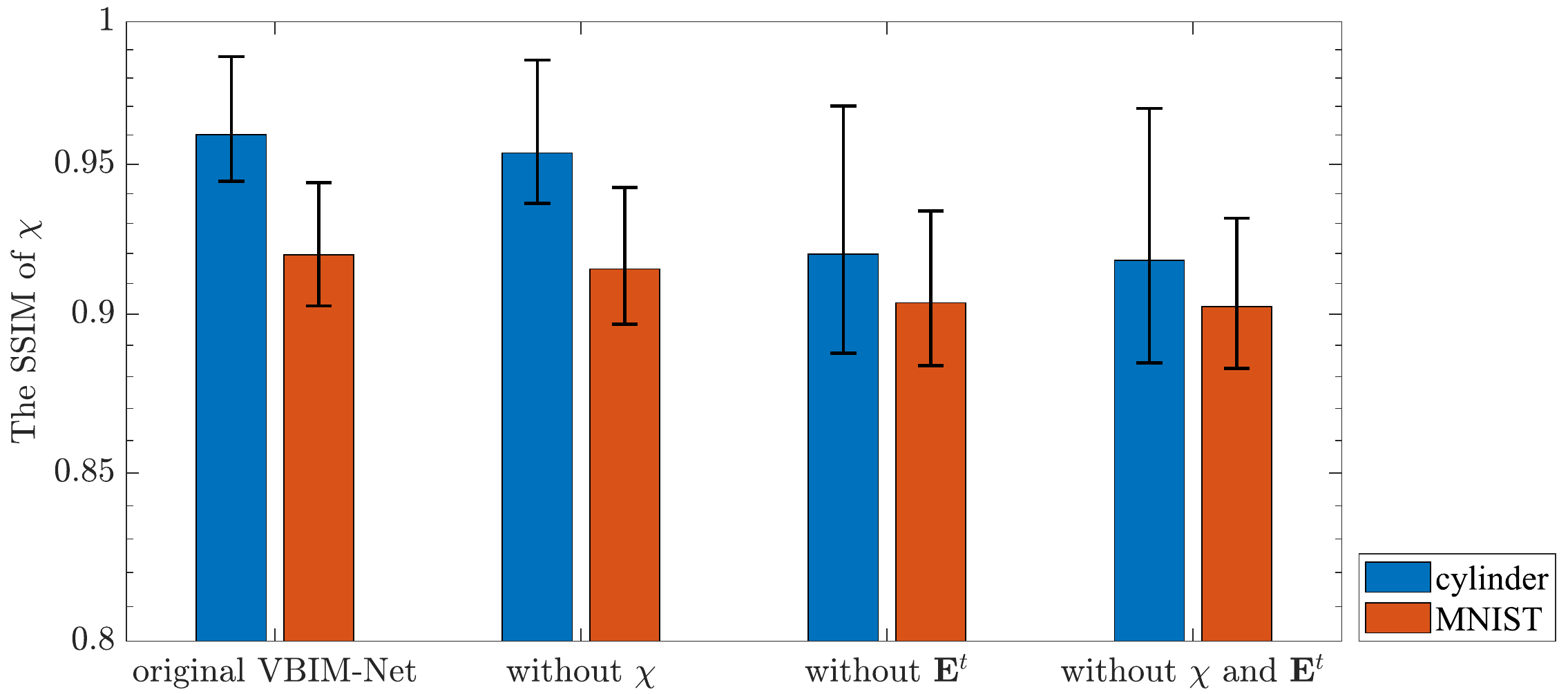}
    \captionsetup{font=footnotesize}
    \caption{The SSIM performance on the test set of VBIM-Net after removing certain input physical quantities. 
    The quartiles Q1 and Q3 are marked.}
    \label{fig:different_input_performance}
\end{figure}

To verify the effectiveness of the subnetwork input configuration in VBIM-Net, 
we trained the following four models with different subnetwork inputs while keeping all other training conditions consistent: 
\begin{enumerate}
    \item The original VBIM-Net, where $\delta\tilde{\boldsymbol{\chi}}_{(k)} \oplus \boldsymbol{\chi}_{(k-1)}$ and $\delta\mathbf{E}^i_{(k)} \oplus \mathbf{E}^{t}_{(k-1)}$ are used as inputs for the $k$-th $\boldsymbol{\chi}$-update and $\mathbf{E}^t$-update networks, respectively; 
    \item Removing the input $\boldsymbol{\chi}_{(k-1)}$ from the $\boldsymbol{\chi}$-update network of VBIM-Net; 
    \item Removing the input $\mathbf{E}^t_{(k-1)}$ from the $\mathbf{E}^t$-update network of VBIM-Net;
    \item Removing both $\boldsymbol{\chi}_{(k-1)}$ and $\mathbf{E}^{t}_{(k-1)}$, keeping only $\delta\tilde{\boldsymbol{\chi}}_{(k)}$ and $\delta\mathbf{E}^i_{(k)}$ as the inputs for the $\boldsymbol{\chi}$-update and $\mathbf{E}^t$-update networks, respectively.
\end{enumerate} 
The SSIM performance of these four models on the test set are shown in Fig. \ref{fig:different_input_performance}, and the first (Q1) and third (Q3) quartiles are marked. 
It can be seen that the original VBIM-Net performs the best, while removing $\boldsymbol{\chi}_{(k-1)}$ or $\mathbf{E}^t_{(k-1)}$ leads to a loss in performance and stability. 
This proves that in VBIM-Net, using the physical quantities output from the previous layer as inputs to the current layer helps the model better learn the updates of these quantities. 

Furthermore, we find that removing $\boldsymbol{\chi}_{(k-1)}$, meaning that only $\delta\tilde{\boldsymbol{\chi}}_{(k)}$ is used as the input for the $\boldsymbol{\chi}$-update network, has a relatively minor impact on model performance, 
which indicates that the approximate contrast variation inspired by VBIM provides sufficient information for VBIM-Net to compute the contrast updates, demonstrating VBIM-Net's advantage in structural rationality. 

\subsubsection{Validation of the Efficacy of the Layer-Wise Constraint}
\begin{figure}[!htbp]
    \centering
    \vspace{-0.2cm}
    \includegraphics[width=0.95\linewidth]{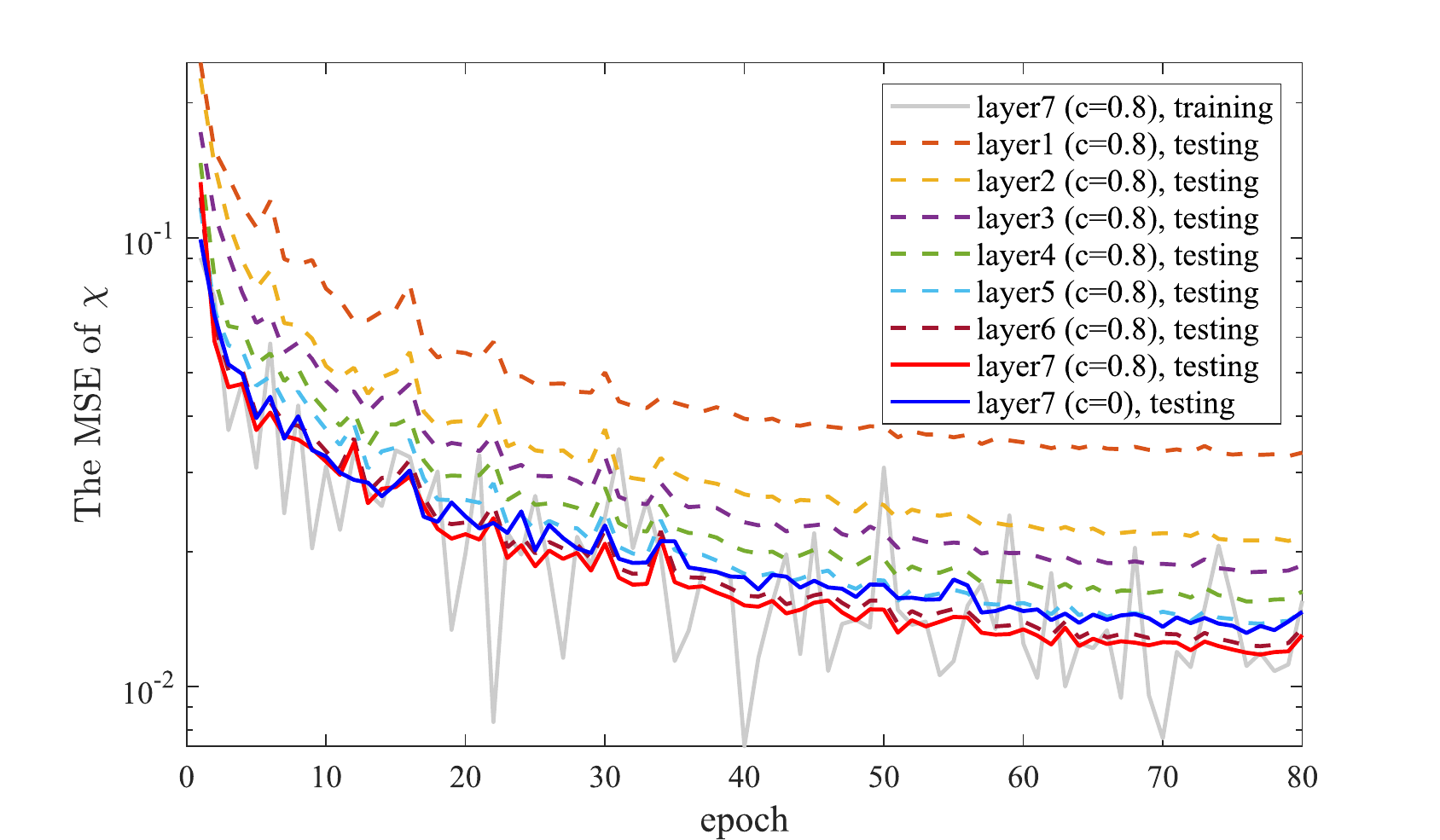}
    \captionsetup{font=footnotesize}
    \caption{The MSE of $\boldsymbol{\chi}$ on the test set with ($c=0.8$) / without ($c=0$) the layer-wise constraint.}
    \label{fig:layer_loss}
\end{figure}

\begin{figure}[!htbp]
    \centering
    \includegraphics[width=\linewidth]{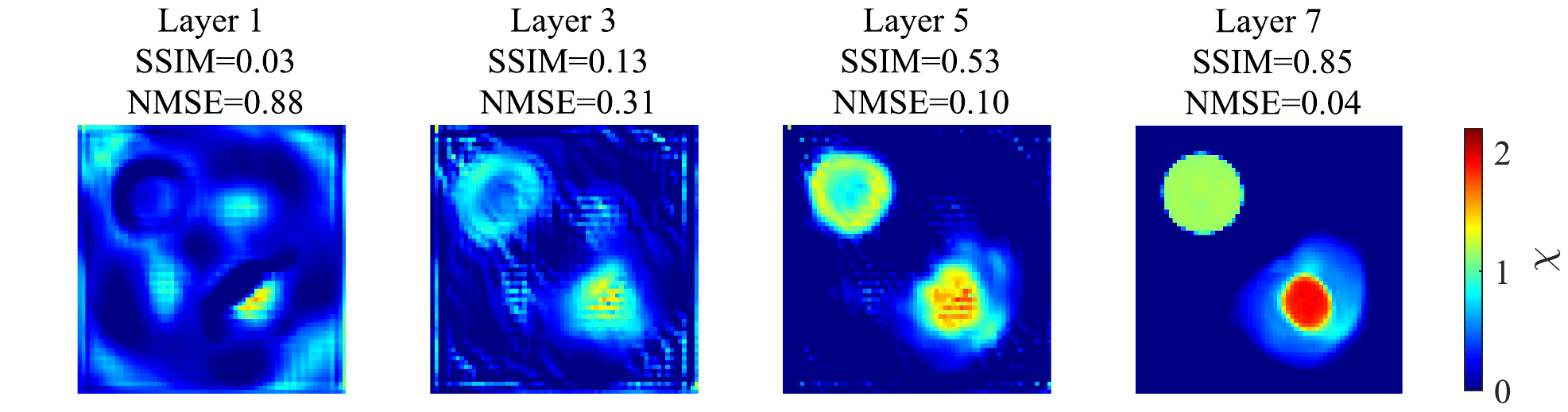}
    \captionsetup{font=footnotesize}
    \caption{The updated reconstructions of VBIM-Net without the layer-wise constraint.}
    \label{fig:recon_layers_no_w}
    \vspace{-0.2cm}
\end{figure}

In order to verify the efficacy of the layer-wise constraint, we perform the following ablation experiment: 
set $c = 0$ in Eq. (\ref{eq: w_k}) to only supervise the output of the last layer of VBIM-Net (where $0^0=1$ is defined), while keeping other settings unchanged. 
Fig. \ref{fig:layer_loss} shows the contrast MSE on the test set during training with/without the layer-wise constraint. 
It also shows the training convergence curve of the final layer output under $c=0.8$. 
When using the layer-wise constraint, the convergence trends of the training set and the test set are essentially consistent, indicating minimal signs of overfitting. 
Additionally, in this case, the reconstruction quality can progressively improve with the stacking of update blocks. 
With the same number of subnetwork layers, the model with the layer-wise constraint can converge to a better performance. 
Fig. \ref{fig:recon_layers_no_w} illustrates the reconstruction results without the layer-wise constraint, where the scatterer is the same as in Fig. \ref{fig:recon_layers_cylinder}. 
It can be seen that removing the layer-wise constraint can result in an ambiguous physical interpretation of the output from intermediate layers, as well as a deterioration in the final reconstruction quality. 

\begin{figure}[!htbp]
    \centering
    \vspace{-0.2cm}
    \includegraphics[width=0.95\linewidth]{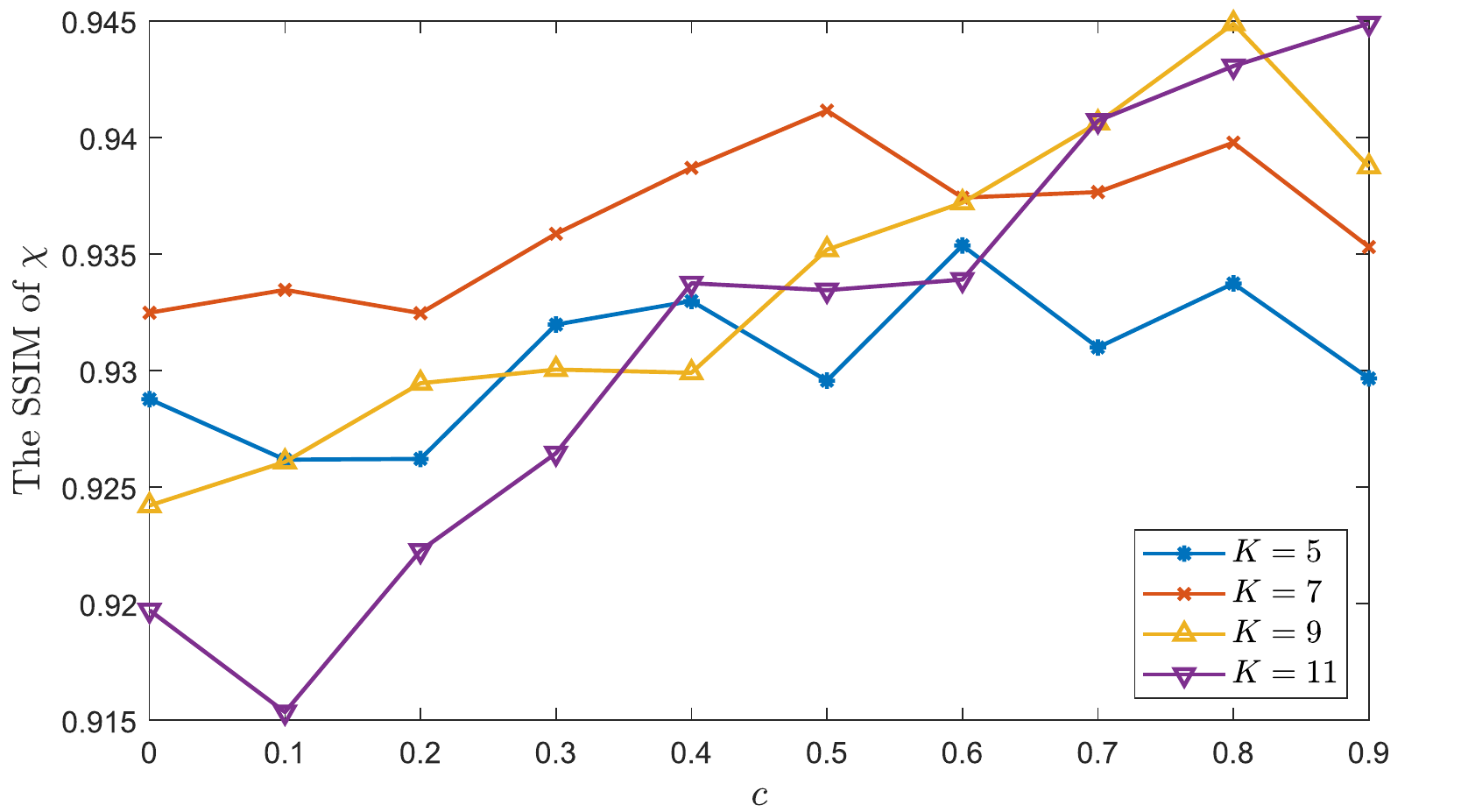}
    \captionsetup{font=footnotesize}
    \caption{The contrast SSIM performance curve of VBIM-Net with different layer numbers $K$, trained under different layer-wise constraint parameters $c$. The performance is evaluated on the test set.}
    \label{fig:performance_c_K_mean}
\end{figure}

\begin{figure*}[!t]
    \centering
    \includegraphics[width=\linewidth]{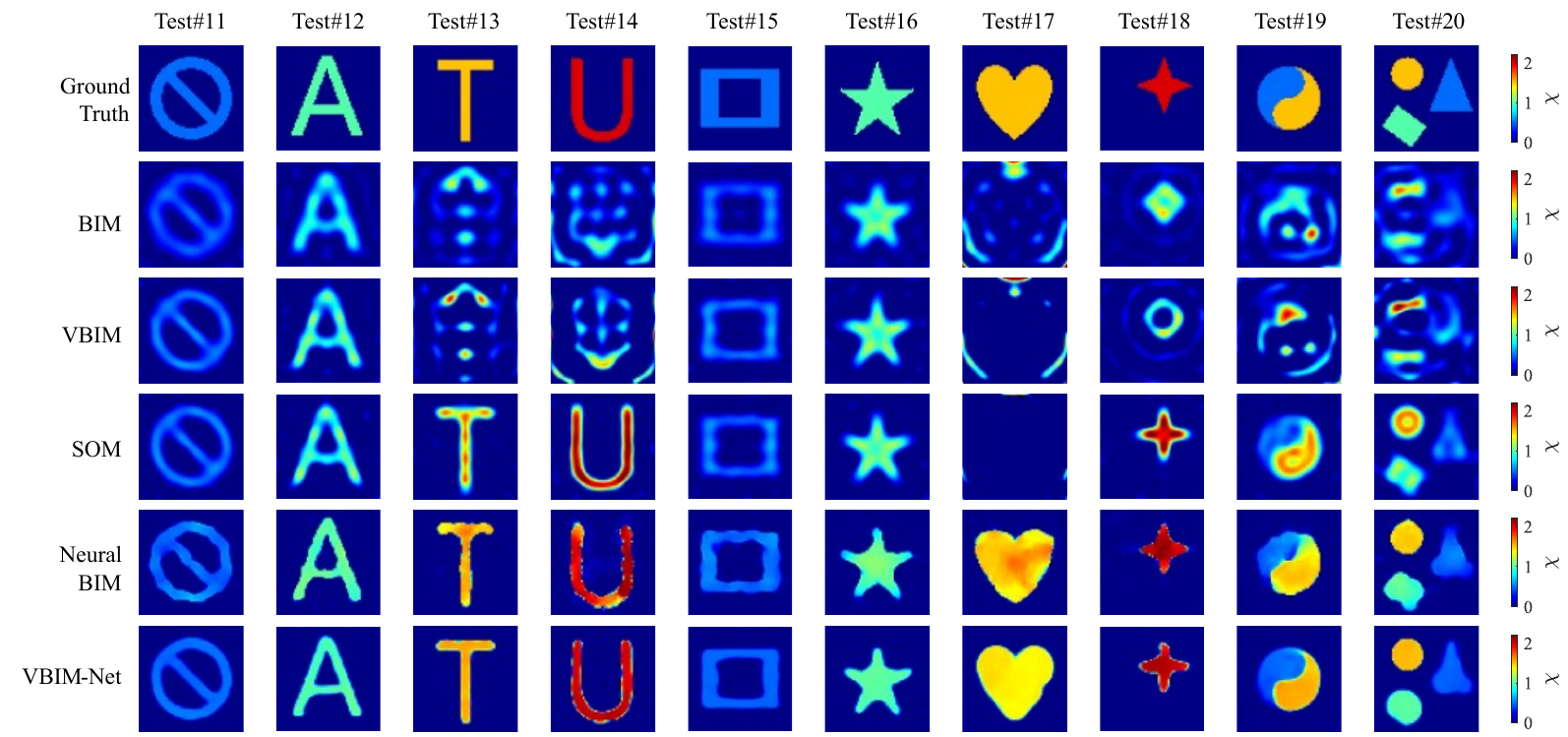}
    \captionsetup{font=footnotesize}
    \caption{Reconstruction results of cross-dataset examples Test\#11-Test\#20 with 10\% Gaussian noise. 
    The contrast of these scatterers is 0.5, 1.0, 1.5, and 2.0.}
    \label{fig:cross_dataset_examples}
\end{figure*}

\begin{table*}[!htbp]
\captionsetup{font=footnotesize}
\caption{Performance Metrics of Reconstruction Results for Test\#11-Test\#20} \label{tab:performance_cross_dataset_example}
\centering
\renewcommand{\arraystretch}{1.1}
\begin{tabular}{|c|c|c|c|c|c|c|c|c|c|c|}
\hline
\multirow{2}{*}{Method} & \multicolumn{2}{c|}{Test\#11} & \multicolumn{2}{c|}{Test\#12} & \multicolumn{2}{c|}{Test\#13} & \multicolumn{2}{c|}{Test\#14} & \multicolumn{2}{c|}{Test\#15}\\
\cline{2-11}
                                 & SSIM & NMSE & SSIM & NMSE & SSIM & NMSE & SSIM & NMSE & SSIM & NMSE \\
\hline
\multicolumn{1}{|c|}{BIM}        & 0.54 & 0.20 & 0.49 & 0.21 & 0.25 & 0.69 & 0.13 & 0.82 & 0.62 & 0.15\\
\hline
\multicolumn{1}{|c|}{VBIM}        & 0.79 & 0.11 & 0.73 & 0.15 & 0.36 & 0.78 & 0.23 & 0.92 & 0.73 & 0.13\\
\hline
\multicolumn{1}{|c|}{SOM}        & 0.81 & 0.10 & 0.80 & 0.13 & 0.83 & 0.12 & 0.76 & \textbf{0.11} & 0.76 & 0.12\\
\hline
\multicolumn{1}{|c|}{NeuralBIM}  & 0.82 & 0.12 & 0.84 & 0.12 & 0.85 & 0.13 & 0.61 & 0.17 & 0.82 & 0.11\\
\hline
\multicolumn{1}{|c|}{VBIM-Net} & \textbf{0.92} & \textbf{0.05} & \textbf{0.93} & \textbf{0.06} & \textbf{0.97} & \textbf{0.03} & \textbf{0.83} & {0.15} & \textbf{0.90} & \textbf{0.07} \\
\hline
\hline
\multirow{2}{*}{Method} & \multicolumn{2}{c|}{Test\#16} & \multicolumn{2}{c|}{Test\#17} & \multicolumn{2}{c|}{Test\#18} & \multicolumn{2}{c|}{Test\#19} & \multicolumn{2}{c|}{Test\#20}\\
\cline{2-11}
                                 & SSIM & NMSE & SSIM & NMSE & SSIM & NMSE & SSIM & NMSE & SSIM & NMSE \\
\hline
\multicolumn{1}{|c|}{BIM}        & 0.55 & 0.15 &  -  &  -  & 0.45 & 0.36 & 0.20 & 0.71 & 0.26 & 0.43\\
\hline
\multicolumn{1}{|c|}{VBIM}        & 0.72 & 0.12 &  -  &  -  & 0.48 & 0.69 & 0.36 & 0.97 & 0.31 & 0.58\\
\hline
\multicolumn{1}{|c|}{SOM}        & 0.77 & 0.12 &  -   &  -   & 0.81 & 0.12 & 0.74 & 0.07 & 0.66 & 0.10\\
\hline
\multicolumn{1}{|c|}{NeuralBIM} & 0.83 & 0.10 & 0.71 & 0.06 & 0.84 & \textbf{0.09} & 0.80 & 0.07 & 0.78 & \textbf{0.07}\\
\hline
\multicolumn{1}{|c|}{VBIM-Net} & \textbf{0.90} & \textbf{0.08} & \textbf{0.88} & \textbf{0.05} & \textbf{0.91} & {0.12} & \textbf{0.93} & \textbf{0.03} & \textbf{0.82} & {0.08} \\
\hline
\end{tabular}
\end{table*}

To further reveal the effect of the layer-wise constraint parameter on the performance of VBIM-Net, 
we plot the contrast SSIM curve of VBIM-Net with different $K$ and $c$ on the test set in Fig. \ref{fig:performance_c_K_mean}. 
When $ K = 5 $, the network has too few layers and lacks the iterative characteristics typically expected.   
In this case, the impact of $ c $ on model performance is not clearly patterned. 
As the number of network layers increases to $K~=~7,~9,~11$, the effect of the layer-wise constraint becomes more evident, and the optimal value for parameter $c$ shows a gradual increase. 
The above observations suggest that the layer-wise constraint is critical for training deep VBIM-Net models. 
In the absence of the layer-wise constraint ($c = 0$), the performance of the model may deteriorate as the model depth $K$ increases. This is attributed to the overfitting caused by the lack of physical constraints. 
Nevertheless, the incorporation of the layer-wise constraint can enhance the stability during training for deep VBIM-Net models, enabling a higher performance upper bound with more update layers. 
Although there is a correlation between the optimal value for parameter $c$ and the number of network layers $K$, in practice, setting $c$ as the default value of 0.8 can achieve satisfactory performance in most cases. 

Since the layer-wise constraint requires intermediate layer outputs and increases the complexity of the loss function, it indeed slightly raises the computational cost of optimizing VBIM-Net. 
According to the experimental results, with the same training configurations, adding the layer-wise constraint increases the training time by $2.5\%$, but it effectively enhances the model's interpretability and scalability. 

The above experiments demonstrate the efficacy of the layer-wise constraint in training deep-unfolding models like VBIM-Net, especially when we need to enhance the model performance through increasing the model's depth. 
This also validates the effectiveness of incorporating soft physical constraints from iterative algorithms into the network design.

\subsection{Synthetic Data Inversion: Cross-Dataset Test} \label{subsec:synthetic_cross}

To verify the generalization ability and robustness of VBIM-Net, 
we test VBIM-Net on more challenging cross-dataset examples, 
evaluate the model's performance under various noise levels, contrast values, and measurement dimensions, 
and extend the model to lossy scatterers.

\subsubsection{Cross-Dataset Samples Reconstruction}
Fig. \ref{fig:cross_dataset_examples} shows the reconstruction results of Test\#11-Test\#20 with 10\% Gaussian noise, 
and the reconstruction quality metrics are listed in Table \ref{tab:performance_cross_dataset_example}. 
It can be seen that VBIM-Net shows superior performance in cases where scatterers have complex shapes and high contrast. 
In the reconstruction of Test\#17, divergence occurred in BIM, VBIM and SOM, whereas VBIM-Net still achieves successful reconstruction. 
Test\#19 tests the splicing of two scatterers with different contrasts, 
and Test\#20 tests three dispersed scatterers with various shapes and contrasts. 
In terms of reconstruction quality of these test samples, VBIM-Net not only outperforms iterative methods such as BIM, VBIM and SOM, but also surpasses the modified NeuralBIM, even with fewer learnable parameters.
These results fully demonstrate the cross-dataset generalization ability of VBIM-Net, 
which is attributed to the physical information introduced in the network architecture and the loss function.
Given the significant performance disadvantage of BIM and VBIM in both within-dataset and cross-dataset tests, we exclude them from the comparisons in the following experiments.

\begin{figure}[!htbp]
    \centering
    % \vspace{-0.2cm}
    \includegraphics[width=0.9\linewidth]{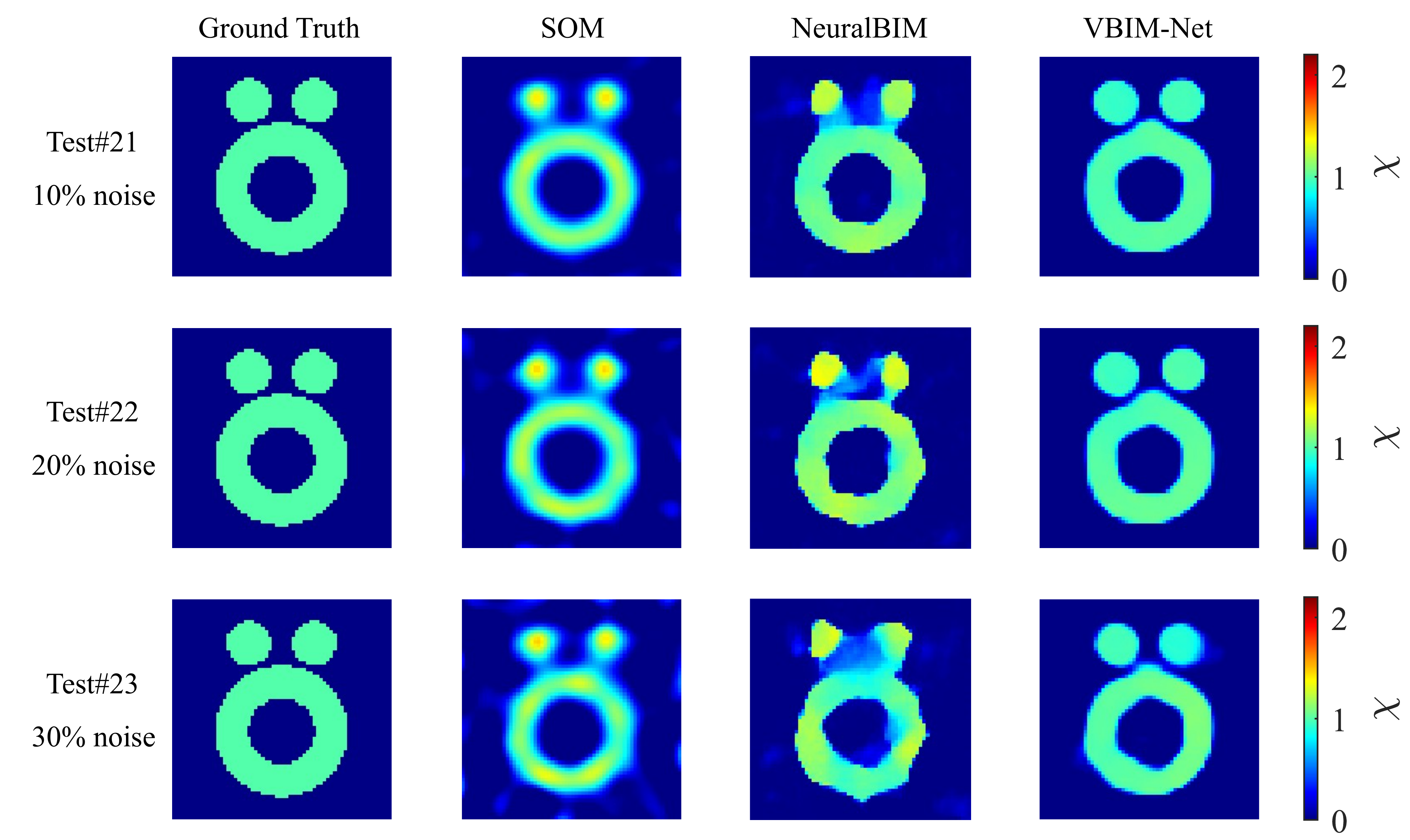}
    \captionsetup{font=footnotesize}
    \caption{Reconstruction results of the ``Austria" profile with contrast $\chi=1.0$ under different noise levels.}
    \label{fig:Austria_diff_nl}
\end{figure}

\begin{table}[!htbp]
\captionsetup{font=footnotesize}
\caption{Performance Metrics of Reconstruction Results for Test\#21-Test\#23} \label{tab:performance_Austria_noise}
\centering
\renewcommand{\arraystretch}{1.1}
\begin{tabular}{|c|c|c|c|c|c|c|}
\hline
\multirow{2}{*}{Method} & \multicolumn{2}{c|}{Test\#21} & \multicolumn{2}{c|}{Test\#22} & \multicolumn{2}{c|}{Test\#23} \\
\cline{2-7}
                                 & SSIM & NMSE & SSIM & NMSE & SSIM & NMSE \\
\hline
\multicolumn{1}{|c|}{SOM}        & 0.69 & 0.11 & 0.66 & 0.12 & 0.65 & 0.13 \\
\hline
\multicolumn{1}{|c|}{NeuralBIM}  & 0.73 & 0.14 & 0.70 & 0.14 & 0.67 & 0.19 \\
\hline
\multicolumn{1}{|c|}{VBIM-Net}   & \textbf{0.91} & \textbf{0.05} & \textbf{0.90} & \textbf{0.06} & \textbf{0.88} & \textbf{0.07} \\
\hline
\end{tabular}
\end{table}

\begin{figure}[!htbp]
    \centering
    \vspace{-0.1cm}
    \includegraphics[width=\linewidth]{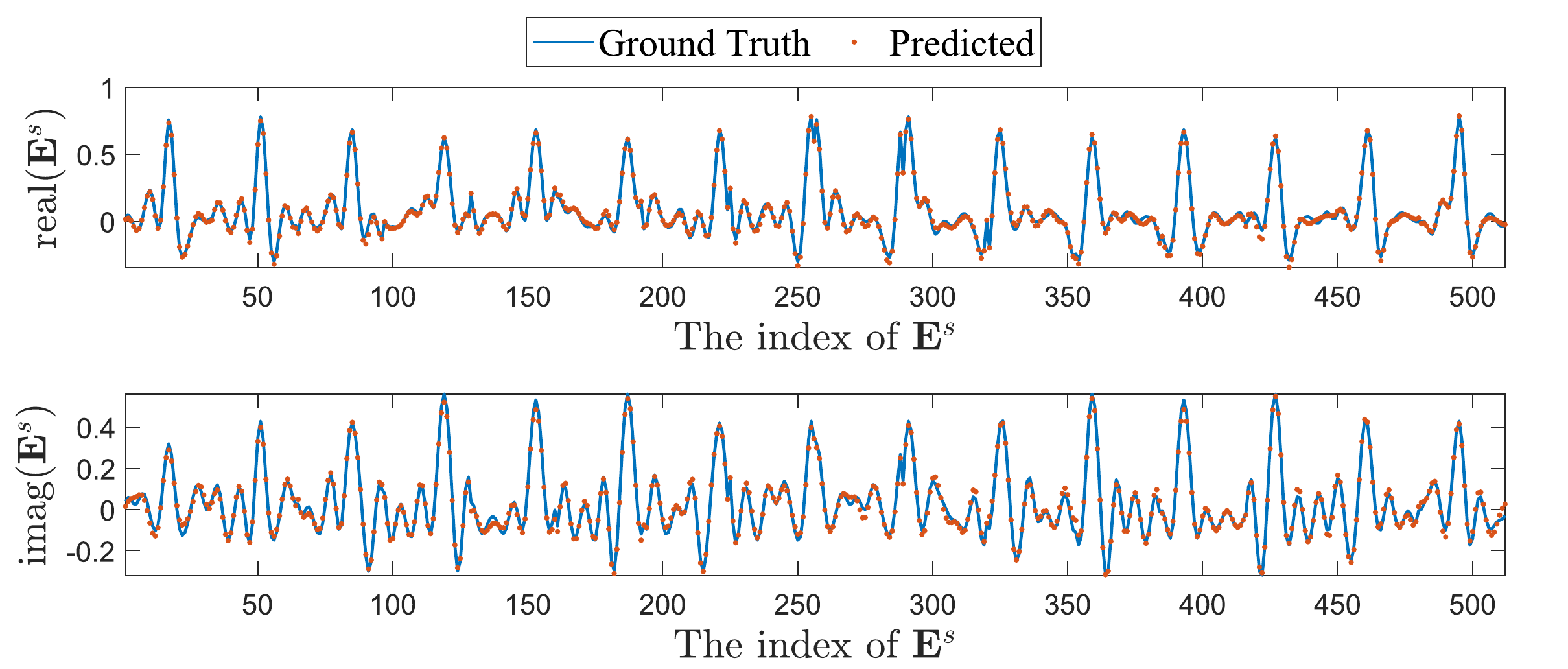}
    \captionsetup{font=footnotesize}
    \caption{The predicted scattered field measurements $\hat{\mathbf{E}}^s$ of Test\#23 with a noise level of 30\%.}
    \label{fig:Es_Austria_1.0_nl30}
\end{figure}

\begin{figure}[!htbp]
    \centering
    \includegraphics[width=\linewidth]{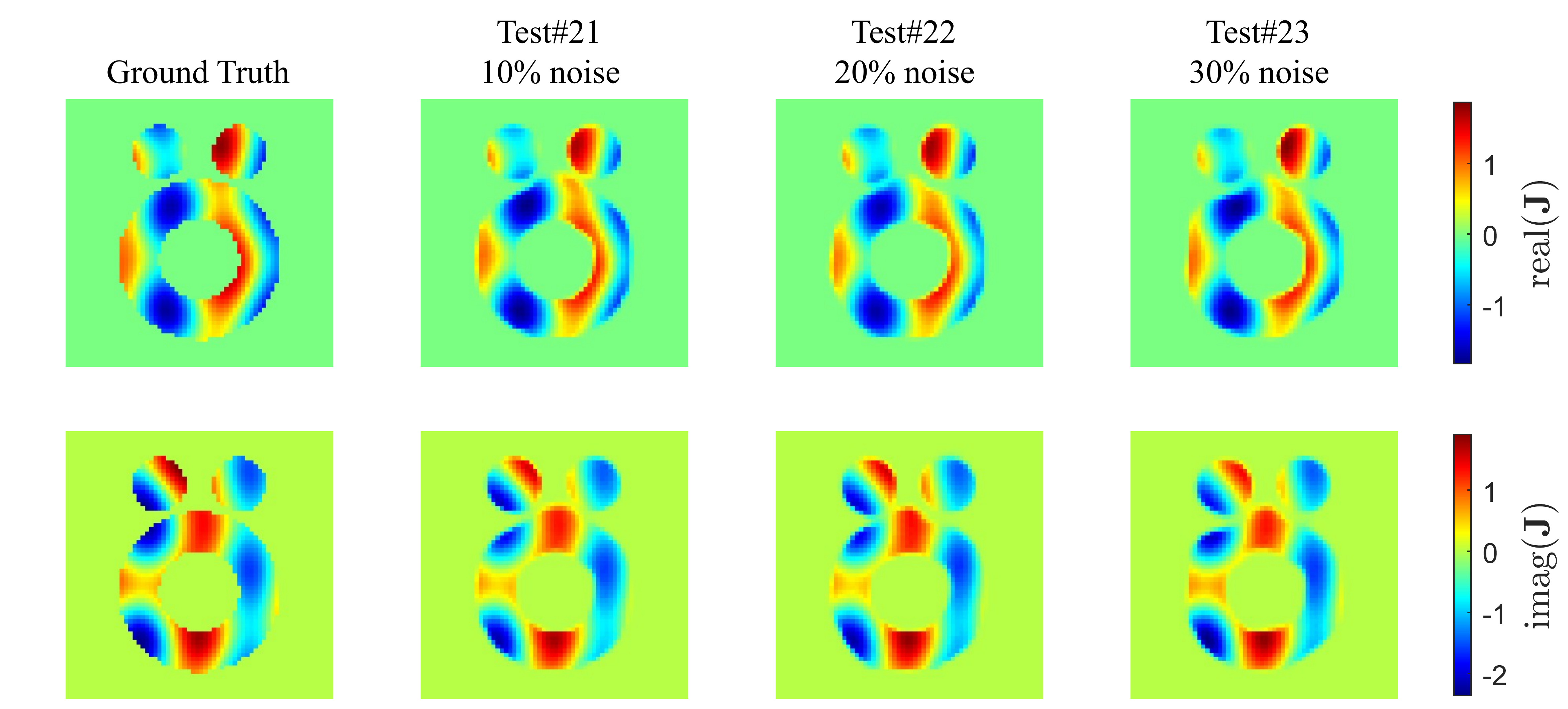}
    \captionsetup{font=footnotesize}
    \caption{The predicted contrast source $\hat{\mathbf{J}}_p$ of Test\#21-Test\#23. 
    The incident wave direction is $0^{\circ}$ ($p = 1$). }
    \label{fig:J_Austria_1.0_diff_nl}
\end{figure}

\subsubsection{Test under Different Noise Levels}
We test the performance of VBIM-Net under different noise levels. 
We choose the ``Austria" profile \cite{belkebir1996Austria} with contrast $\chi=1.0$ as the target scatterer 
and test the reconstruction performance when adding 10\% (SNR=20dB), 20\% (SNR=14dB), and 30\% (SNR=10dB) Gaussian noise. 
The reconstruction results of Test\#21-Test\#23 under different noise levels are shown in Fig. \ref{fig:Austria_diff_nl}, and their corresponding performance metrics are listed in Table \ref{tab:performance_Austria_noise}. 
Experimental results show that VBIM-Net can achieve better reconstruction quality than other two methods under various noise levels. 
This is mainly due to VBIM-Net's physics-inspired iterative framework and the training scheme with extra noise, which enhance the network's robustness. 

From the reconstructed contrast $\hat{\boldsymbol{\chi}}$ and total field $\hat{\mathbf{E}}^t$, we can derive the predicted scattered field measurements, 
\begin{equation}
    \hat{\mathbf{E}}^s = \mathbf{G}_{S} \operatorname{diag}\left(\hat{\boldsymbol{\chi}}\right) \hat{\mathbf{E}}^{t}. 
\end{equation}
Fig. \ref{fig:Es_Austria_1.0_nl30} shows the predicted $\hat{\mathbf{E}}^s$ of VBIM-Net when the noise level is 30\%.  
It can be seen that VBIM-Net can accurately predict scattered field measurements under high noise levels, which further proves VBIM-Net's robustness against noise.
Fig. \ref{fig:J_Austria_1.0_diff_nl} shows the predicted contrast source for the ``Austria'' profile under various noise levels when the incidence direction is $0^{\circ}$. 
The predicted contrast source $\hat{\mathbf{J}}_p$ for the $p$th incidence is depicted as
\begin{equation}
    \hat{\mathbf{J}}_p = \operatorname{diag}\left(\hat{\boldsymbol{\chi}}\right)\hat{\mathbf{E}}^{t}_p. 
\end{equation}
It can be seen that VBIM-Net can maintain the consistency of various physical quantities that are not supervised during training, 
such as contrast current and scattered field, with their actual values while reconstructing contrast images, 
which indicates the physical interpretability of VBIM-Net.

\subsubsection{Test with Different Contrast}

\begin{figure}[!htbp]
    \centering
    \includegraphics[width=\linewidth]{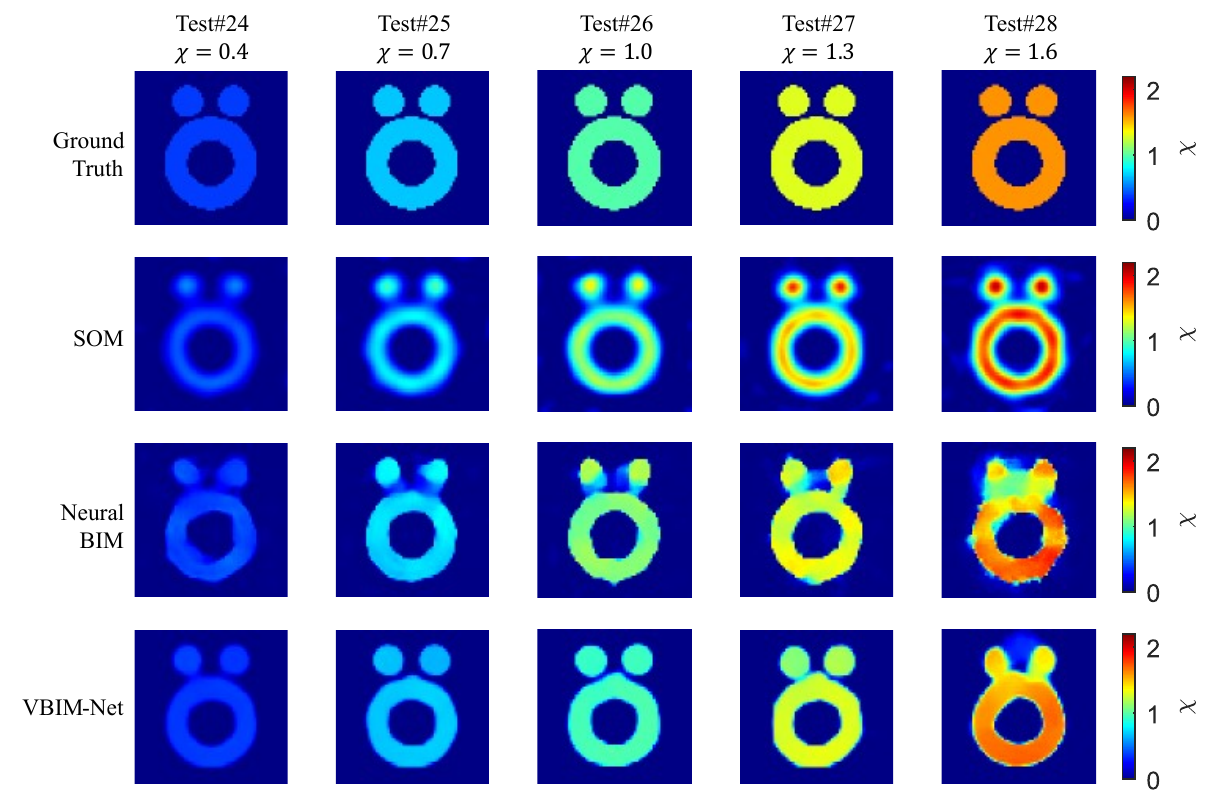}
    \captionsetup{font=footnotesize}
    \caption{Reconstruction results of the ``Austria'' profile under 10\% noise level. 
    The contrast is equal to 0.4, 0.7, 1.0, 1.3, and 1.6, respectively.}
    \label{fig:Austria_diff_contrast}
\end{figure}

\begin{figure*}[!htbp]
    \centering
    \hfill
    \begin{subfigure}[t]{0.48\linewidth}
        \includegraphics[width=\linewidth]{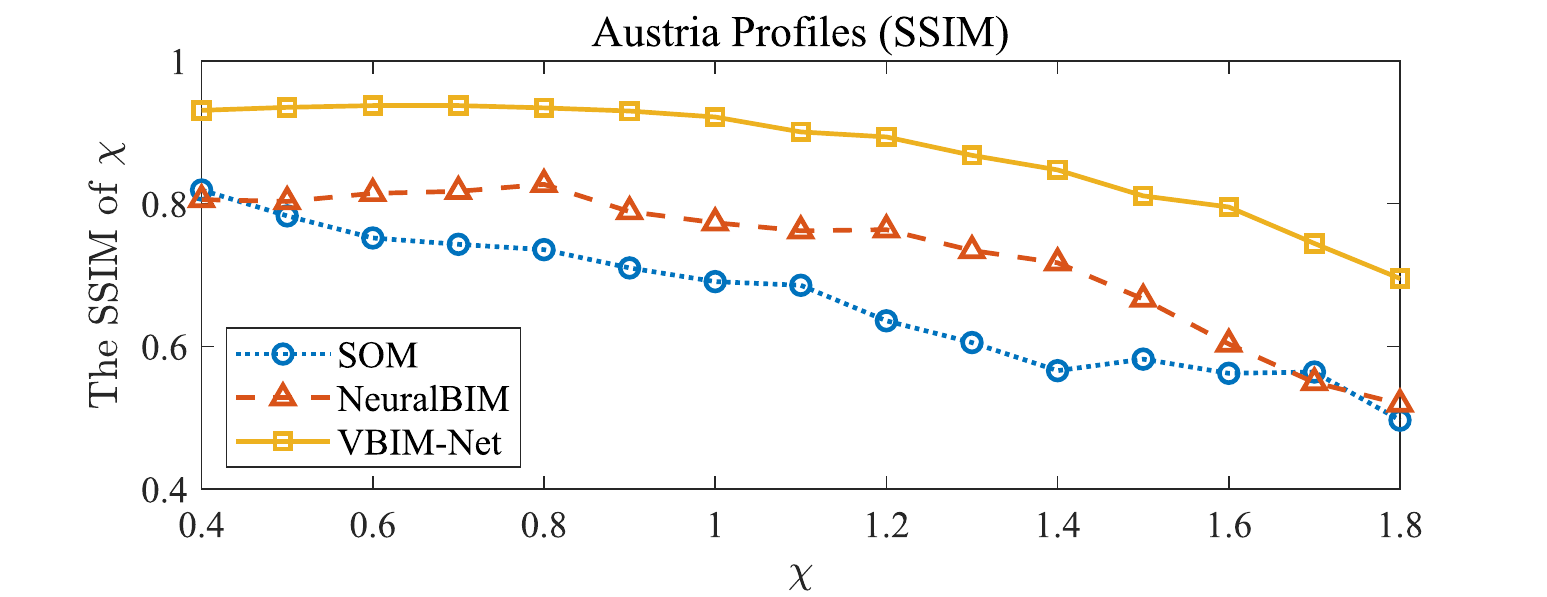}
        \label{fig:Austria_x_ssim}
    \end{subfigure}
    \hfill
    \begin{subfigure}[t]{0.48\linewidth}
        \includegraphics[width=\linewidth]{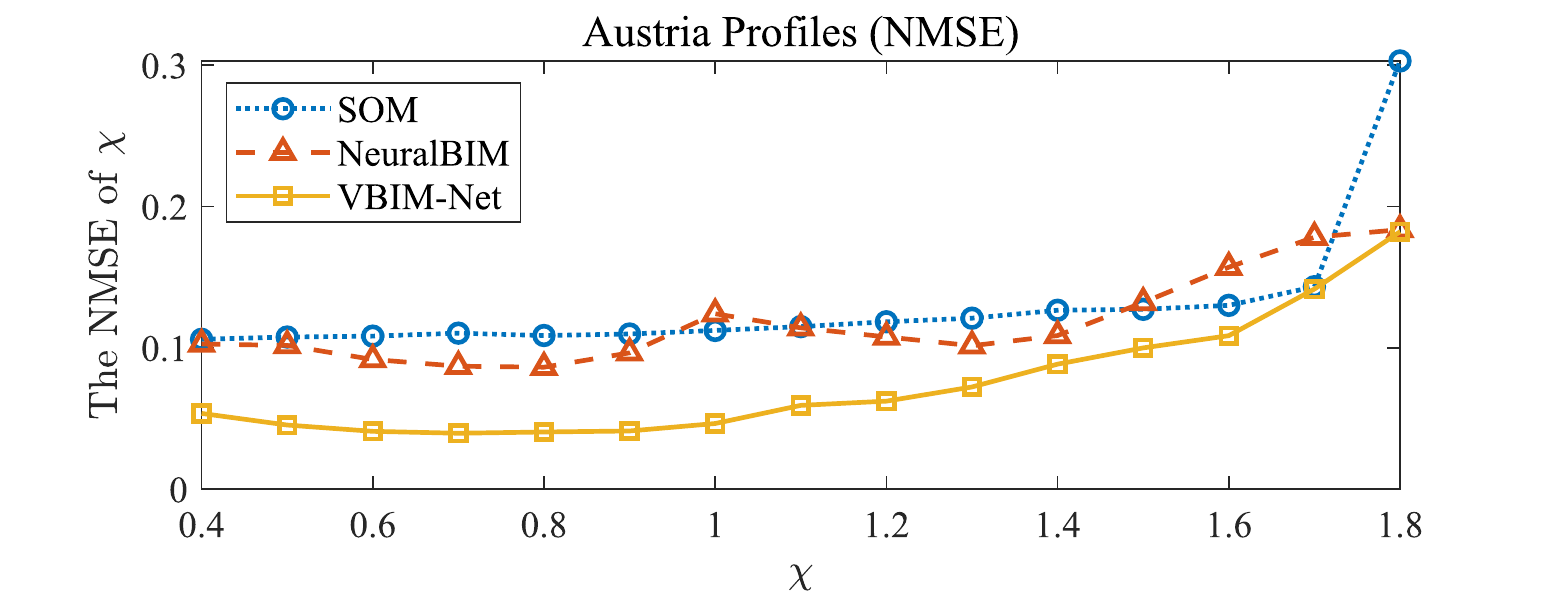}
        \label{fig:Austria_x_nmse}
    \end{subfigure}
    \hfill
    \vspace{-0.45cm}
    \captionsetup{font=footnotesize}
    \caption{Reconstruction performance of the ``Austria'' profile with different contrast by SOM, NeuralBIM, and VBIM-Net. 
    The noise level is 10\%. }
    \vspace{-0.2cm}
    \label{fig:Austria_x_ssim_nmse}
\end{figure*}

Furthermore, we investigate the influence of scatterer contrast on the inversion performance of VBIM-Net. 
Fig. \ref{fig:Austria_diff_contrast} illustrates the reconstruction results for the ``Austria'' profile with 10\% Gaussian noise when the contrast is 0.4, 0.7, 1.0, 1.3, and 1.6.
Fig. \ref{fig:Austria_x_ssim_nmse} illustrates the inversion performance of SOM, the modified NeuralBIM, and the proposed VBIM-Net for the "Austria" profile with contrast from 0.4 to 1.8. 
We find that VBIM-Net exhibits competitive inversion performance for scatterers with various contrast. 
However, as the contrast increases, the performance of these three methods will deteriorate significantly. 
Even though we have included high-contrast cases in the training set, 
the inversion for high-contrast scatterers is still a challenge for VBIM-Net 
because it is extended from the Born iteration framework, which can only operate well in low-contrast situations. 

\begin{figure*}[!t]
    \centering
    % \vspace{-0.3cm}
    \hfill
    \begin{subfigure}[b]{0.48\linewidth}
        \includegraphics[width=\linewidth]{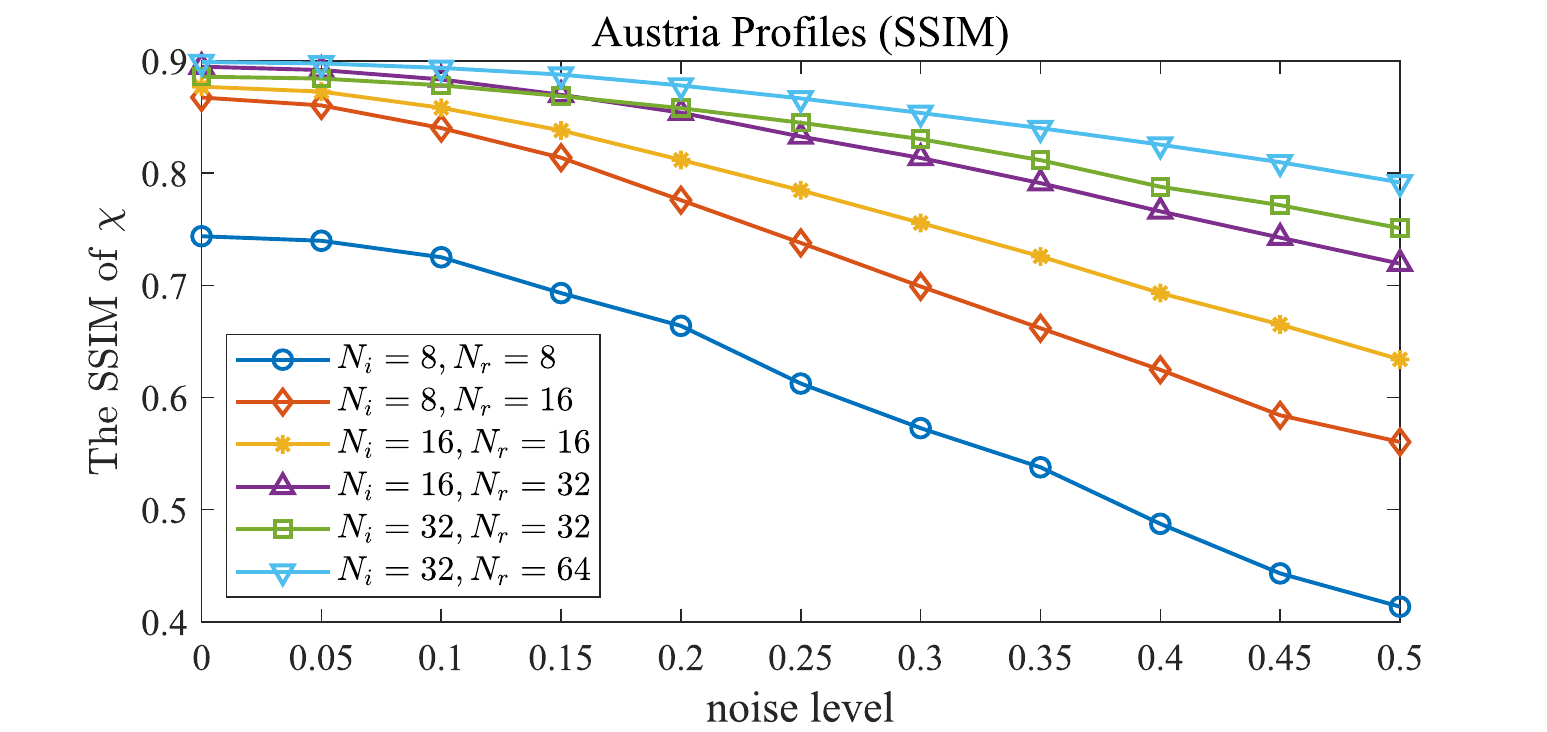}
        \label{fig:curve_ssim}
    \end{subfigure}
    \hfill
    \begin{subfigure}[b]{0.48\linewidth}
        \includegraphics[width=\linewidth]{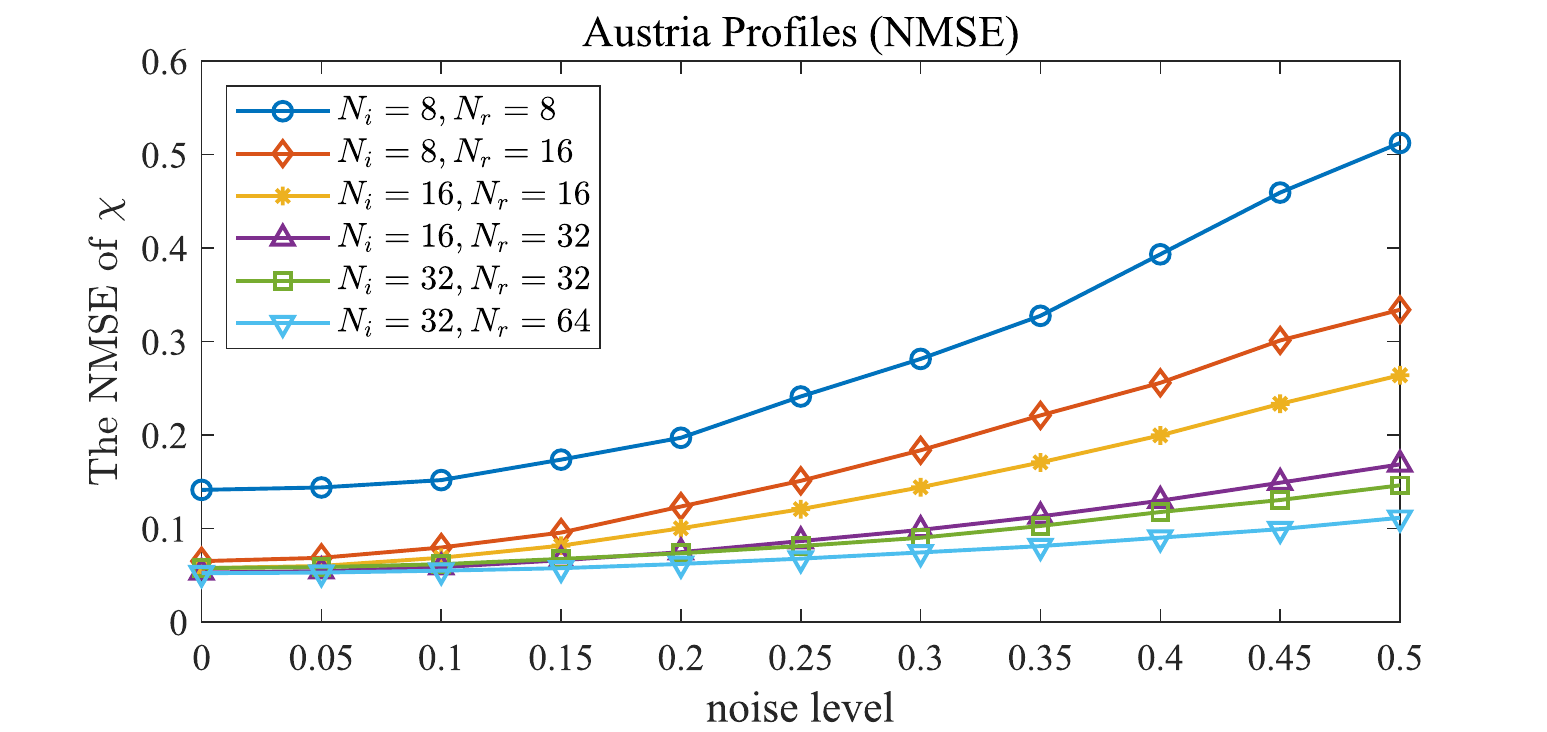}
        \label{fig:curve_nmse}
    \end{subfigure}
    \hfill
    \vspace{-0.45cm}
    \captionsetup{font=footnotesize}
    \caption{Reconstruction performance of the ``Austria'' profile under different scattered field measurement dimensions and different noise levels by VBIM-Net.
    The scatterer is the ``Austria" profile with contrast $\chi = 1.0$. }
    \label{fig:curve_snr_antenna}
\end{figure*}

\begin{figure*}[!t]
    \centering
    \includegraphics[width=\linewidth]{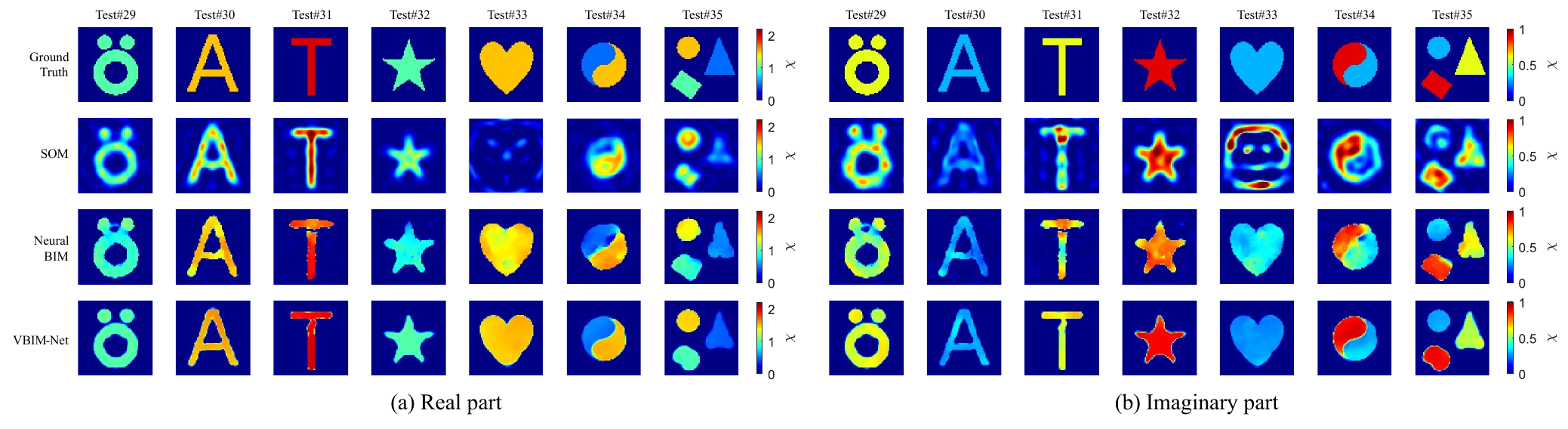}
    \captionsetup{font=footnotesize}
    \caption{Reconstruction results of lossy scatterers Test\#29-Test\#35 with 10\% Gaussian noise.
    The real parts of the contrast of these scatterers are 0.5, 1.0, 1.5, 2.0 and the imaginary parts are 0.3, 0.6, 0.9. }
    \label{fig:lossy_dataset_examples}
\end{figure*}

\begin{table*}[!htbp]
\captionsetup{font=footnotesize}
% \vspace{0.5cm}
\caption{Performance Metrics of the Real Part of the Reconstruction Results for Test\#29-Test\#35} \label{tab:performance_lossy_dataset_example_real}
\centering
\renewcommand{\arraystretch}{1.1}
\begin{tabular}{|c|c|c|c|c|c|c|c|c|c|c|c|c|c|c|}
\hline
\multirow{2}{*}{Method} & \multicolumn{2}{c|}{Test\#29} & \multicolumn{2}{c|}{Test\#30} & \multicolumn{2}{c|}{Test\#31} & \multicolumn{2}{c|}{Test\#32} & \multicolumn{2}{c|}{Test\#33} & \multicolumn{2}{c|}{Test\#34} & \multicolumn{2}{c|}{Test\#35}\\
\cline{2-15}
                                 & SSIM & NMSE & SSIM & NMSE & SSIM & NMSE & SSIM & NMSE & SSIM & NMSE & SSIM & NMSE & SSIM & NMSE \\
\hline
\multicolumn{1}{|c|}{SOM}        & 0.38 & 0.15 & 0.31 & 0.19 & 0.33 & 0.16 & 0.49 & 0.16 &   -  &  -   & 0.31 & 0.11 & 0.37 & 0.14\\
\hline
\multicolumn{1}{|c|}{NeuralBIM}  & 0.82 & 0.10 & 0.81 & 0.12 & 0.82 & 0.15 & 0.78 & 0.16 & 0.81 & 0.05 & 0.82 & 0.08 & 0.77 & 0.09\\
\hline
\multicolumn{1}{|c|}{VBIM-Net} & \textbf{0.93} & \textbf{0.04} & \textbf{0.93} & \textbf{0.06} & \textbf{0.96} & \textbf{0.05} & \textbf{0.90} & \textbf{0.08} & \textbf{0.93} & \textbf{0.02} & \textbf{0.96} & \textbf{0.02} & \textbf{0.87} & \textbf{0.06} \\
\hline
\end{tabular}
\end{table*}

\begin{table*}[!htbp]
\captionsetup{font=footnotesize}
\caption{Performance Metrics of the Imaginary Part of the Reconstruction Results for Test\#29-Test\#35} \label{tab:performance_lossy_dataset_example_imag}
\centering
\renewcommand{\arraystretch}{1.1}
\begin{tabular}{|c|c|c|c|c|c|c|c|c|c|c|c|c|c|c|}
\hline
\multirow{2}{*}{Method} & \multicolumn{2}{c|}{Test\#29} & \multicolumn{2}{c|}{Test\#30} & \multicolumn{2}{c|}{Test\#31} & \multicolumn{2}{c|}{Test\#32} & \multicolumn{2}{c|}{Test\#33} & \multicolumn{2}{c|}{Test\#34} & \multicolumn{2}{c|}{Test\#35}\\
\cline{2-15}
                                 & SSIM & NMSE & SSIM & NMSE & SSIM & NMSE & SSIM & NMSE & SSIM & NMSE & SSIM & NMSE & SSIM & NMSE \\
\hline
\multicolumn{1}{|c|}{SOM}        & 0.27 & 0.20 & 0.16 & 0.57 & 0.15 & 0.56 & 0.26 & 0.18 &  -   &  -   & 0.16 & 0.19 & 0.21 & 0.20\\
\hline
\multicolumn{1}{|c|}{NeuralBIM}  & 0.69 & 0.10 & 0.69 & 0.15 & 0.62 & 0.21 & 0.51 & 0.16 & 0.76 & 0.14 & 0.61 & 0.13 & 0.66 & 0.12\\
\hline
\multicolumn{1}{|c|}{VBIM-Net} & \textbf{0.93} & \textbf{0.04} & \textbf{0.94} & \textbf{0.06} & \textbf{0.96} & \textbf{0.06} & \textbf{0.88} & \textbf{0.08} & \textbf{0.94} & \textbf{0.04} & \textbf{0.95} & \textbf{0.02} & \textbf{0.84} & \textbf{0.10} \\
\hline
\end{tabular}
\end{table*}

\begin{figure}[!htbp]
    \centering
    \hfill
    \begin{subfigure}[b]{0.46\linewidth}
        \includegraphics[width=\linewidth]{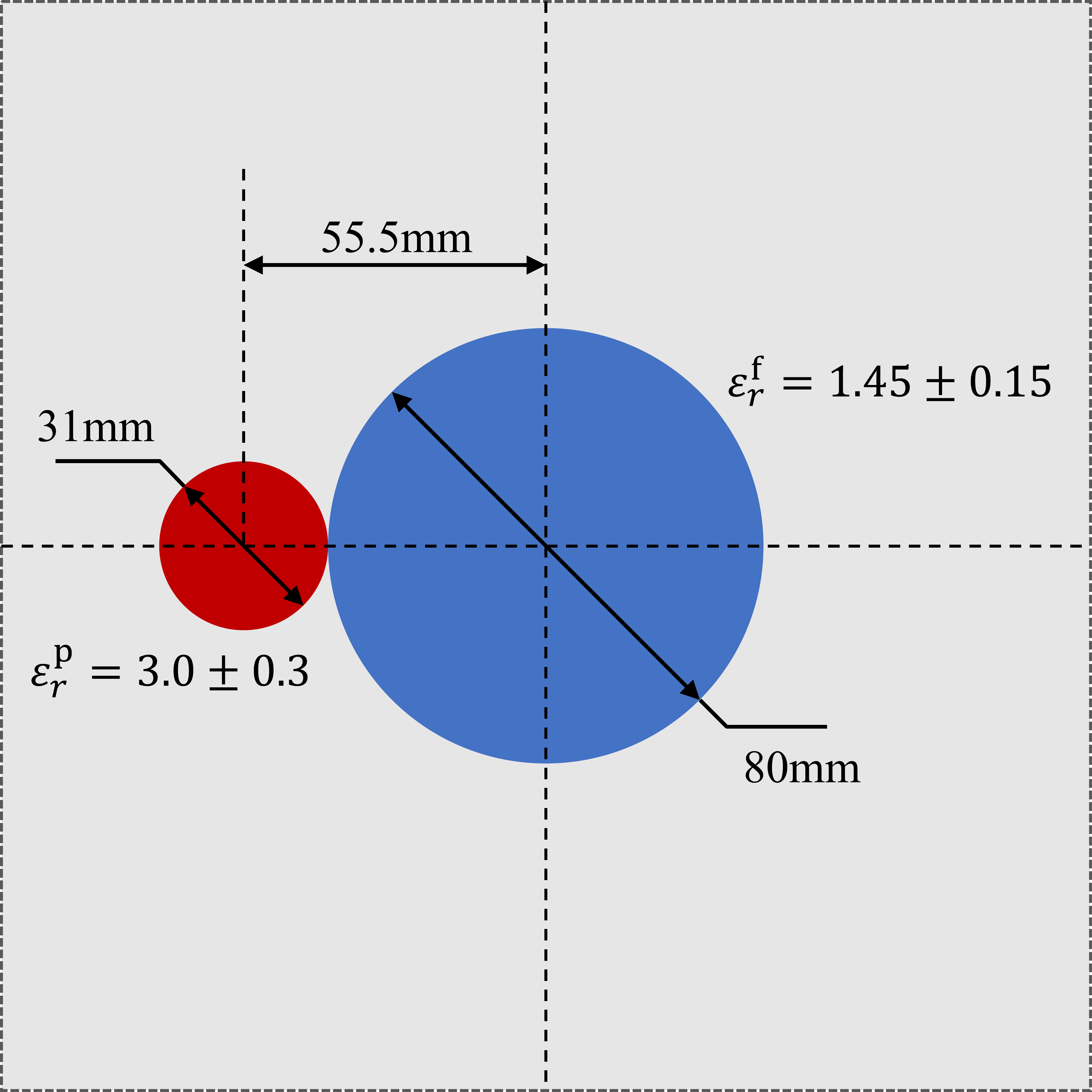}
        \captionsetup{font=footnotesize}
        \caption{``FoamDielExt" profile.}
        \label{fig:FoamDielExt}
    \end{subfigure}
    \hfill
    \begin{subfigure}[b]{0.46\linewidth}
        \includegraphics[width=\linewidth]{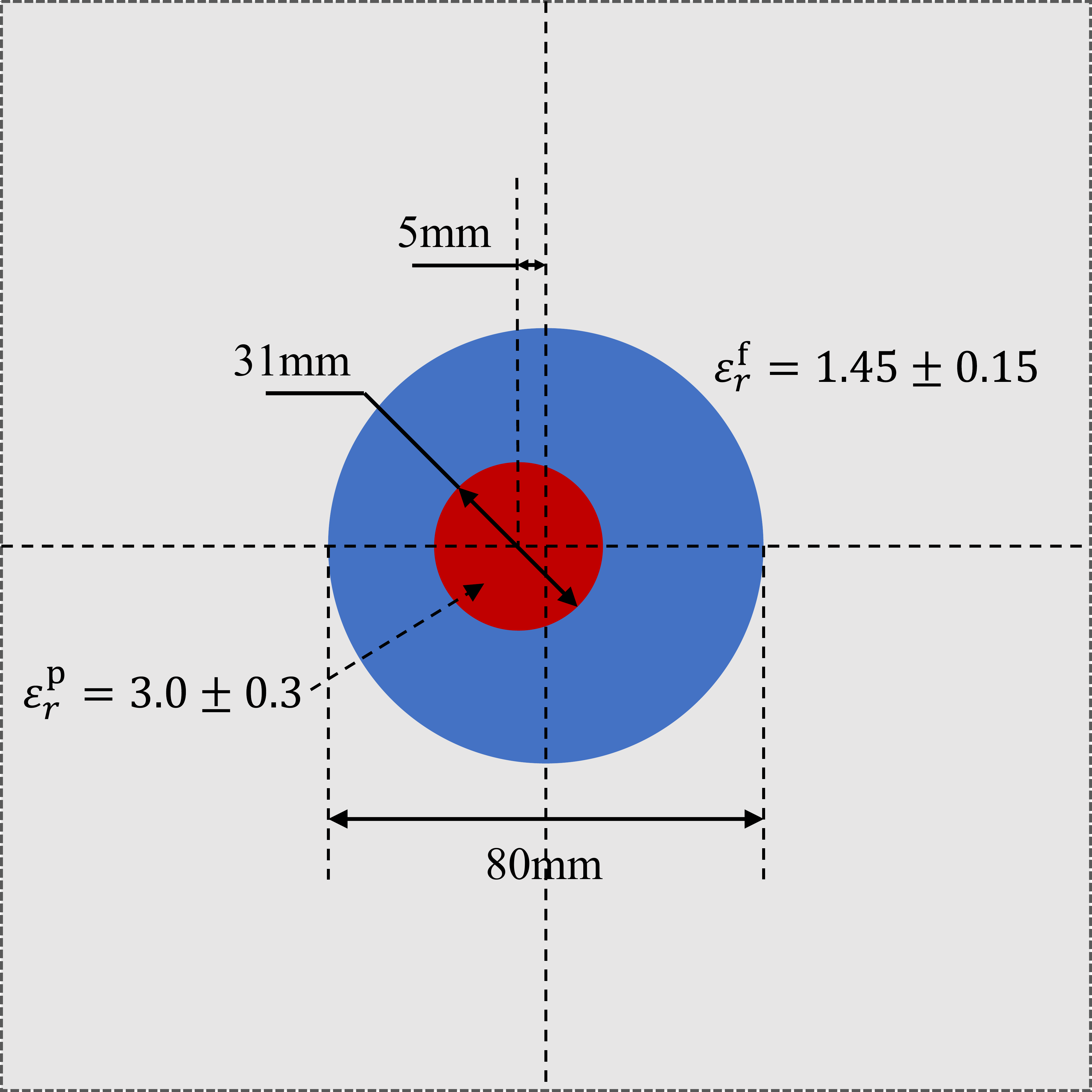}
        \captionsetup{font=footnotesize}
        \caption{``FoamDielInt" profile.}
        \label{fig:FoamDielInt}
    \end{subfigure}
    \hfill
    \captionsetup{font=footnotesize}
    \caption{``FoamDielExt" and ``FoamDielInt" profiles from the Fresnel experimental data.}
    \label{fig:FoamDiel}
\end{figure}

\subsubsection{Test with Different Measurement Dimensions}
Fig. \ref{fig:curve_snr_antenna} evaluates the reconstruction performance of VBIM-Net across various noise levels and measurement dimensions, 
where the scatterer is fixed as the ``Austria" profile with contrast $\chi = 1.0$. 
Despite VBIM-Net being trained within a noise range of 5\%-35\%, it exhibits acceptable performance even under higher noise levels during testing. 
This resilience can be attributed to the loss weighting strategy, which is proportional to SNR, enabling the model to adapt its reconstruction approach under variational noise levels. 
According to the chart, the performance degradation from the case $N_i=8, N_r=16$ to the case $N_i=8, N_r=8$ is apparent. 
This is because the sparse receiving antennas is difficult to precisely characterize the complete scattered field with measurement data, thus intensifying the ill-posedness of the ISP. 
Meanwhile, collecting more scattered field measurement data not only improves the upper bound of reconstruction performance but also significantly enhances the model's robustness against noise.

\subsubsection{Test with Lossy Scatterers}
Finally, we test VBIM-Net's reconstruction ability for lossy scatterers. 
For the VBIM-Net model, we only need to modify the input and output channel number of contrast variation update network $\mathcal{F}^{\delta\boldsymbol{\chi}}_{k}$, keeping other model parameters and hyperparameters unchanged. 
We switch to using the lossy scatterer dataset for training models. The real and imaginary parts of the reconstructed contrast for Test\#29-Test\#35 are shown in Fig. \ref{fig:lossy_dataset_examples}, 
and their corresponding quality metrics are summarized in Tables \ref{tab:performance_lossy_dataset_example_real} and \ref{tab:performance_lossy_dataset_example_imag}. 
It demonstrates that VBIM-Net can perform high-quality reconstruction of diverse lossy scatterers. 
Compared with the lossless situation, VBIM-Net exhibits more remarkable performance advantages than traditional iterative algorithms such as SOM. 
These experimental results further verify the effectiveness of VBIM-Net in addressing ISPs.

\subsection{Experimental Data Inversion} \label{subsec:experimental}

In this section, we test VBIM-Net on experimental data from the Fresnel Institute \cite{Fresnel_2005}. 
We considered the ``FoamDielExt" profile and the ``FoamDielInt" profile, as shown in Fig. \ref{fig:FoamDiel}. 
They consist of a foam cylinder with a diameter of 80mm and a relative permittivity $\varepsilon_r^{\mathrm{f}}=1.45\pm 0.15$, 
and a plastic cylinder with a diameter of 31mm and a relative permittivity $\varepsilon_r^{\mathrm{p}}=3.0\pm 0.3$. 
Different from the previous configuration of synthetic data, 
there are 8 TXs and 241 RXs located on a circle with a radius of 1.67m, 
and the RXs are evenly distributed in the angular range of $60^\circ$-$300^\circ$ relative to each TX. 
We select the measurement data with operating frequencies of 3 GHz and 4 GHz, 
keeping the DOI size at $0.2\mathrm{m}\times 0.2\mathrm{m}$ and the inversion grid resolution at $64\times 64$. 
We use the lossless dataset with cylindrical scatterers mentioned in Section \ref{subsec:implementation_details} as training data. 
The model configurations are consistent with the default settings of the synthetic data inversion experiments. 

\begin{figure}[!t]
    \centering
    \begin{subfigure}[t]{\linewidth}
        \includegraphics[width=0.99\linewidth]{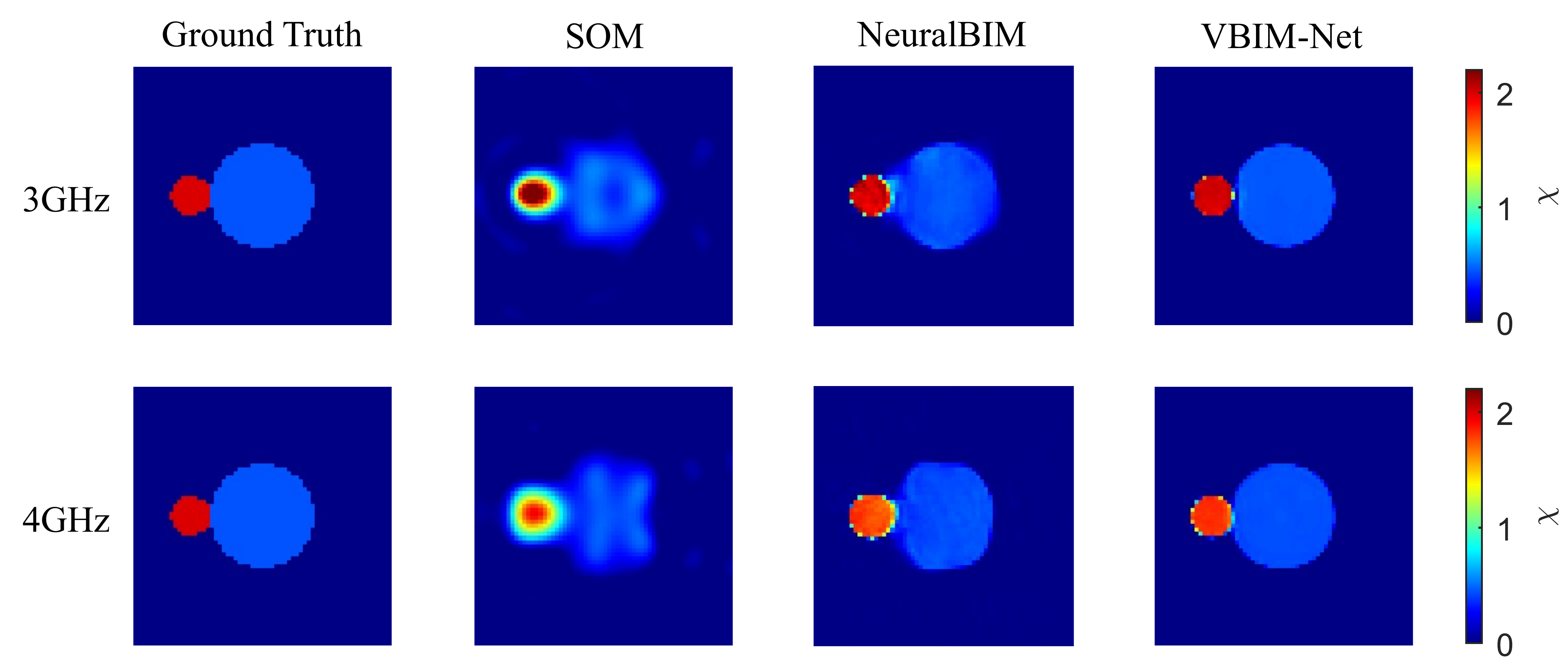}
        \captionsetup{font=footnotesize}
        \caption{``FoamDielExt" profile.}
        % \vspace{2pt}
        \vspace{0.2cm}
        \label{fig:FoamDielExt_3G_4G}
    \end{subfigure}
    % \vspace{0.4cm}
    \begin{subfigure}[t]{\linewidth}
        \includegraphics[width=0.99\linewidth]{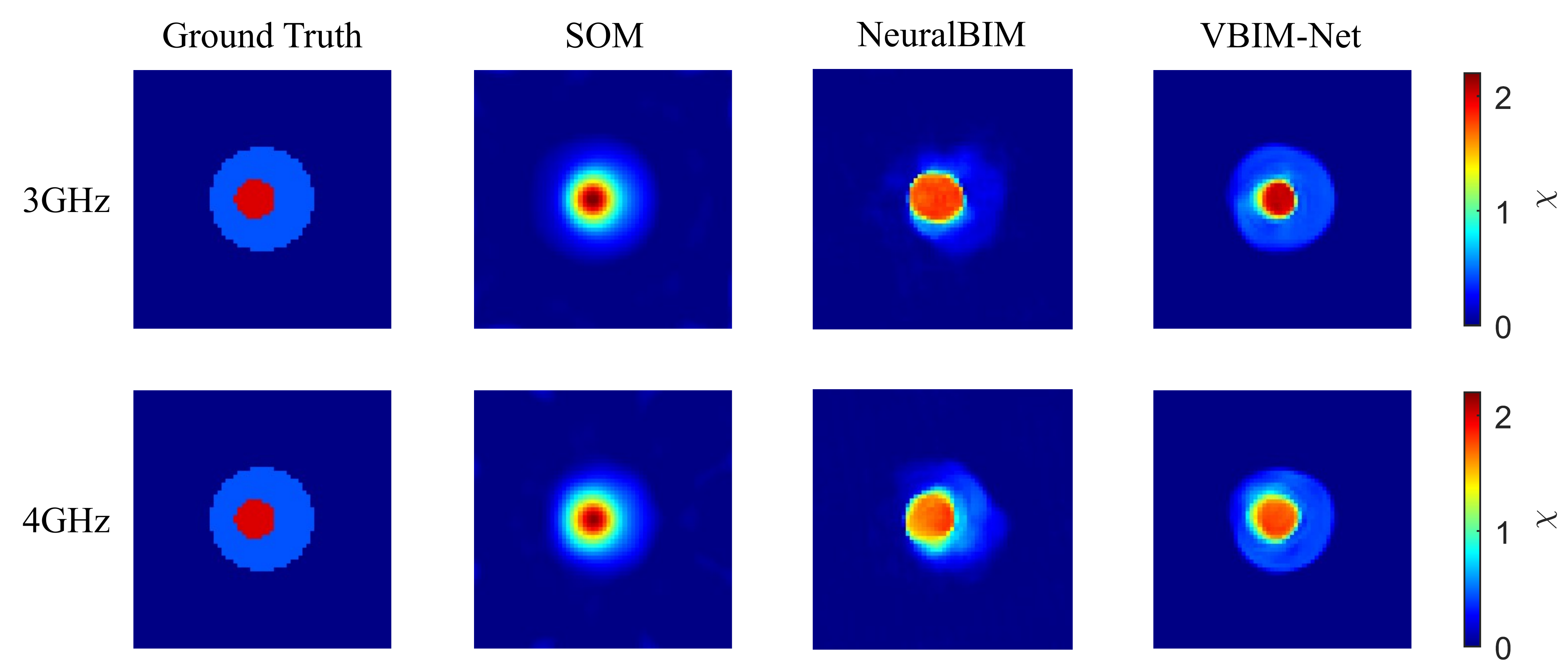}
        \captionsetup{font=footnotesize}
        \caption{``FoamDielInt" profile.}
        \label{fig:FoamDielInt_3G_4G}
    \end{subfigure}
    \captionsetup{font=footnotesize}
    \caption{Reconstruction Results of ``FoamDielExt" and ``FoamDielInt" profiles at 3 GHz and 4 GHz.}
    \label{fig:FoamDiel_3G_4G}
\end{figure}

\begin{figure}[!t]
    \centering
    \begin{subfigure}[t]{0.99\linewidth}
        \includegraphics[width=\linewidth]{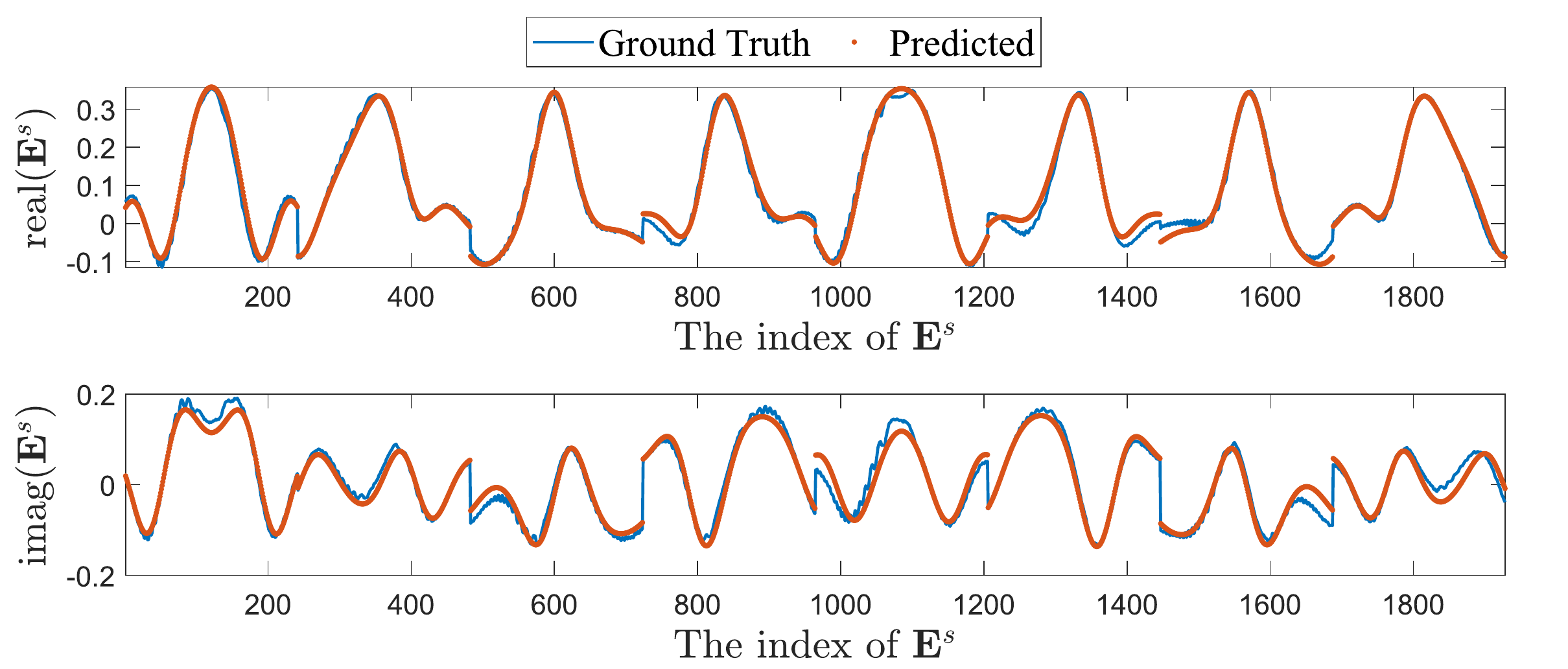}
        \captionsetup{font=footnotesize}
        \caption{$f = 3 \, \mathrm{GHz}$.}
        \vspace{0.2cm}
        \label{fig:Es_FoamDielExtTM_3GHz}
    \end{subfigure}
    \begin{subfigure}[t]{0.99\linewidth}
        \includegraphics[width=\linewidth]{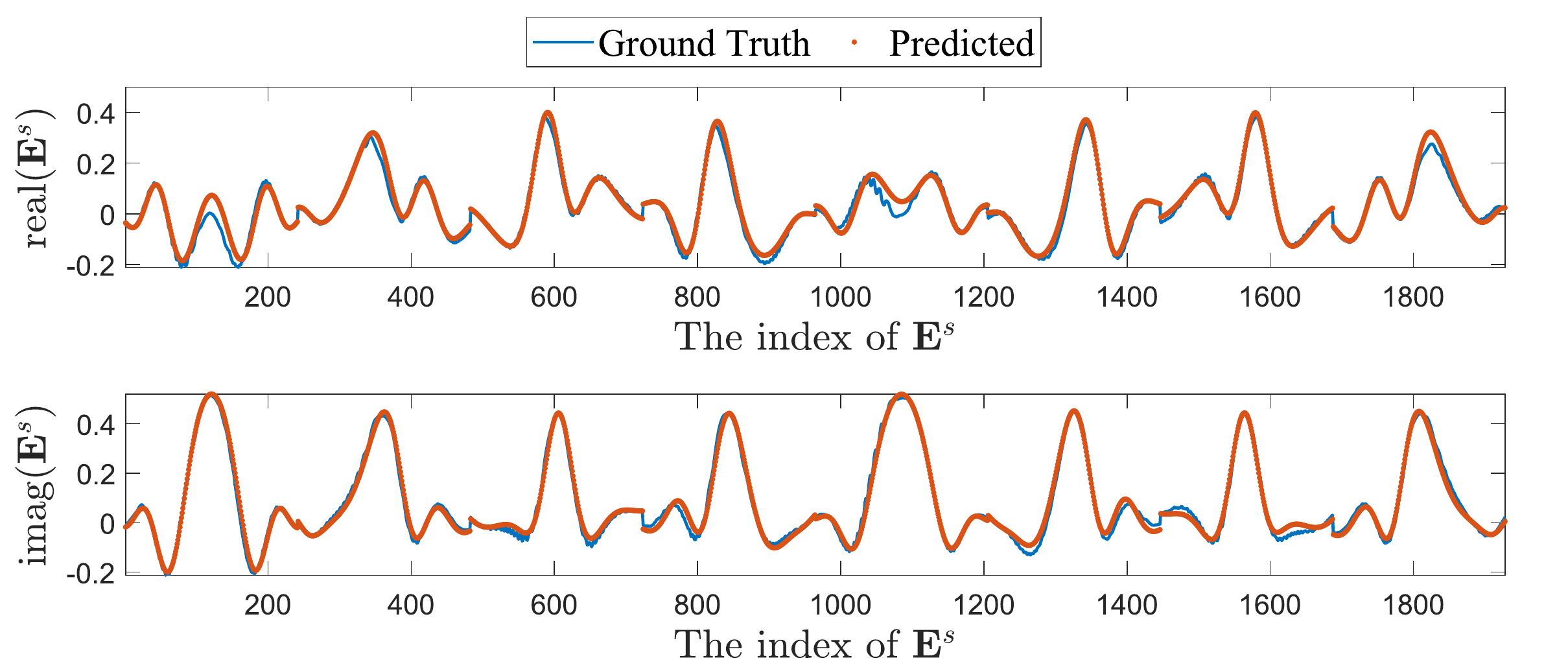}
        \captionsetup{font=footnotesize}
        \caption{$f = 4 \, \mathrm{GHz}$.}
        \label{fig:Es_FoamDielExtTM_4GHz}
    \end{subfigure}
    \captionsetup{font=footnotesize}
    \caption{The prediction of the scattered field measurements of the ``FoamDielExt'' profile at 3 GHz and 4 GHz.}
    \label{fig:Es_FoamDielExtTM}
\end{figure}

Fig. \ref{fig:FoamDiel_3G_4G} shows the reconstruction results of ``FoamDielExt'' and ``FoamDielInt'' profiles at 3 GHz and 4 GHz, and the corresponding quality metrics are summarized in Table \ref{tab:FoamDiel_3G_4G}. 
Fig. \ref{fig:Es_FoamDielExtTM} shows the predicted scattered field of VBIM-Net for the ``FoamDielExt'' profile. 
The results based on experimental data demonstrate that VBIM-Net achieves preferable contrast reconstructions and accurate scattered field predictions at different frequencies, which further verify the effectiveness of the proposed VBIM-Net. 

\begin{table*}[!t]
\captionsetup{font=footnotesize}
\vspace{0.1cm}
\caption{Performance Metrics of the Reconstruction Results of ``FoamDielExt" and ``FoamDielInt" Profiles at 3 GHz and 4 GHz} \label{tab:FoamDiel_3G_4G}
\centering
\begin{tabular}{|c|c|c|c|c|c|c|c|c|}
\hline
{Profile} & \multicolumn{4}{c|}{FoamDielExt} & \multicolumn{4}{c|}{FoamDielInt} \\
\hline
{$f$/GHz} & \multicolumn{2}{c|}{3.0} & \multicolumn{2}{c|}{4.0}  & \multicolumn{2}{c|}{3.0} & \multicolumn{2}{c|}{4.0}\\
\hline
{Method}     & SSIM & NMSE & SSIM & NMSE & SSIM & NMSE & SSIM & NMSE\\
\hline
{SOM}        & 0.81 & 0.16 & 0.80 & 0.21 & 0.78 & 0.13 & 0.81 & 0.14 \\
\hline
{NeuralBIM} & 0.94 & \textbf{0.04} & 0.88 & 0.12 & 0.76 & 0.26 & 0.80 & 0.19 \\
\hline
{VBIM-Net} & \textbf{0.98} & \textbf{0.04} & \textbf{0.98} & \textbf{0.05}  & \textbf{0.95} & \textbf{0.06} & \textbf{0.95} & \textbf{0.10}\\
\hline
\end{tabular}
\end{table*}

\section{Conclusion} \label{sec:conclusion}
In this paper, we propose VBIM-Net, a field-type DL scheme designed to solve full-wave ISPs by emulating the iterative process of VBIM through multiple layers of subnetworks. 
Inspired by the VBIM algorithm, we incorporate an analytical computation in the contrast update step, which converts the scattered field residual into an approximate contrast variation and then refines it using a U-Net. 
This approach avoids the need for matched measurement dimensions and grid resolution required in NeuralBIM. 
To ensure the consistency between predicted and actual physical quantities within each subnetwork, 
we introduce a layer-wise constraint in the loss function, which serves as a soft physical constraint. 
This constraint enforces the data flow of VBIM-Net comply with an iterative process, thereby enhancing the reliability and generalization of deep models. 
In the training stage, we adopt a noisy training scheme to further improve the model's robustness. 
Testing results on synthetic and experimental data verify the superior performance of VBIM-Net compared to BIM, SOM and NeuralBIM, 
and a series of experiments also prove the significant effect of the new design we introduced. 
Overall, VBIM-Net achieves high reconstruction quality, strong generalization ability, and good physical interpretability, providing new ideas for the design of field-type DL schemes. 
Despite these advancements, the proposed VBIM-Net still can be further explored in some aspects, 
such as semi-supervised and unsupervised learning schemes, fully decoupling the model from the measurement settings, and flexible reconstruction resolutions.

\bibliographystyle{IEEEtran}
\bibliography{ref}

\end{document}